\documentclass[nofootinbib,prd,twocolumn,showpacs,showkeys,preprintnumbers]{revtex4-1}
\usepackage{hyperref,amssymb,amsmath,mathrsfs,bm,graphicx}
\begin{document}

\title {Cracking of  general relativistic anisotropic polytropes}
\author{L. Herrera}
\email{lherrera@usal.es}
\affiliation{Escuela de F\'\i sica, Facultad de Ciencias, Universidad Central de Venezuela, Caracas 1050, Venezuela and Instituto Universitario de F\'isica
Fundamental y Matem\'aticas, Universidad de Salamanca 37007, Salamanca, Spain}
\author{E. Fuenmayor}
\email{efuenma@ciens.ucv.ve}
\affiliation{Escuela de F\'\i sica, Facultad de Ciencias, Universidad Central de Venezuela, Caracas 1050, Venezuela}
\author{P. Le\'on}
\email{pabloleon360@gmail.com}
\affiliation{Escuela de F\'\i sica, Facultad de Ciencias, Universidad Central de Venezuela, Caracas 1050, Venezuela}
\begin{abstract}
We discuss the effect that small  fluctuations of local anisotropy of pressure, and of energy density, may have on the occurrence of cracking in spherical
compact objects, satisfying a polytropic equation of state. Two different kinds of polytropes are
considered. For both, it is shown that departures from equilibrium may
lead to the appearance of cracking,  for a wide range of values of the parameters defining the polytrope. Prospective applications of the obtained results, to some astrophysical scenarios, are pointed out.
\end{abstract}
\maketitle
\section{Introduction}
In recent papers the general formalism to study  polytropes for anisotropic matter has been presented, both in the Newtonian  \cite{1p} and the general relativistic regimes  \cite{2p, 3p}. The motivations to undertake such a task were, on the one hand, the  fact that the polytropic equations of state  allow  to deal with a variety of fundamental astrophysical problems (see Refs. \cite{cha, sch,  2, 3, 7', 9, 8,10, 4a, 4b, 4c, 5, 6, 7, 11, 12, 13, pol1n} and references therein).  

 On the other hand,  the local anisotropy of pressure may be caused by a large variety of physical phenomena of the kind  we expect to find  in compact objects (see Refs. \cite{14, 04, hmo02, hod08, hsw08, p1, p2, anis1, anis2, anis3, anis4bis}, and references therein  for an extensive discussion on this point). 

Among all possible sources of anisotropy (see \cite{14} for a comprehensive discussion on this point),  let us mention two  which might be  particularly related to our primary interest. The first one is the intense magnetic field observed in compact objects such as white dwarfs,  neutron stars, or magnetized strange quark stars  (see, for example, Refs. \cite{15, 16, 17, 18, 19} and references therein). Indeed, it is a well-established fact  that  a magnetic field acting on a Fermi gas produces pressure anisotropy (see  Refs. \cite{ 23, 24, 25, 26} and references therein). In some way, the magnetic field can be addressed as a fluid anisotropy (however, it should be observed that for the thermodynamic pressure of degenerate particles under the conditions of Landau quantization, it remains isotropic \cite{bla}). 

 Another source of anisotropy expected to be present  in neutron stars and, in general, in highly dense matter,  is the viscosity (see \cite{Anderson, sad, alford, blaschke, drago, jones, vandalen, Dong} and references therein). At this point it is worth noticing that we are not concerned  by how small the resulting anisotropy produced by the viscosity might be, since as we shall see below, the occurrence of crackings  may happen even for slight deviations from  isotropy.

An alternative approach to anisotropy comes from kinetic theory
using the spherically symmetric Einstein-Vlasov equations, which admits a very rich class of static solutions, none of them isotropic (Ref. \cite{ref1, andreasson, ref2} and
references therein). The advantages or disadvantages of either approach being related to the specific problem under consideration.

The theory of polytropes is based on the polytropic equation of state, which in the Newtonian case reads
\begin{equation}P=K\rho_0^{\gamma}=K\rho_0^{1+1/n} ,\label{Pol}\end{equation}
where $P$ and $\rho_0$ denote the isotropic pressure and the  mass (baryonic) density,  respectively. Constants $K$, $\gamma$, and $n$ are usually called  the polytropic constant, polytropic exponent, and polytropic index, respectively.

In the general relativistic, anisotropic, case, two  possible extensions of the above equation of state are possible, namely:

\begin{enumerate}
\item \begin{equation}P_r=K\rho_0^{\gamma}=K\rho_0^{1+1/n} \label{p1},\end{equation} 
\item \begin{equation}P_r=K\rho^{\gamma}=K\rho^{1+1/n} \label{p2},\end{equation} 
\end{enumerate}
where $P_r$  and $\rho$ denote the radial pressure (see below), and the energy density respectively.

As it should be expected, the assumption of either (\ref{p1}) or (\ref{p2}) is not enough to integrate completely the  field equations, since the appearance of  two principal stresses (instead of one as in the isotropic case) leads to  a system of two equations for three unknown functions.

Thus  in  order to integrate the obtained system of equations we need to provide further information about the anisotropy, inherent to the problem under consideration. For doing that, in \cite{2p}   a particular  ansatz  \cite{anis4} has been assumed, which  allows for specific modelling. This  method  links the obtained  models continually with the isotropic case, thereby allowing us to bring out the influence of anisotropy on the structure
of the object, even for very small anistropies. Here we shall adopt the above mentioned ansatz. However it should be stressed that our models are  presented with the sole  purpose of illustrating the occurrence of crackings; the natural way to obtain models, consists of providing the specific information about the kind of anisotropy present in each specific problem.

The concept of cracking was introduced to describe the
behavior of a fluid distribution just after its departure from equilibrium,
when total nonvanishing radial forces of different signs appear within the
system \cite{cn}.

Thus, if we take a snapshot of the system, just after leaving  equilibrium (on a time scale smaller than the hydrostatic time scale), we say that there is a cracking whenever this radial force is directed
inwards in the inner part of the sphere and reverses its sign beyond some
value of the radial coordinate. In the opposite case, when the force is
directed outwards in the inner part and changes sign in the outer part, we
say that there is an overturning. Further developments on this issue may be found in \cite{cnbis, cn1, cn2, cn4, cn5, cn7bis, cn7bbis, mardam}. As should be clear at this point, the concept of cracking is closely related to  the problem of structure formation \cite{cn3, cn6, cn7}.

In the example examined in \cite{cn} it appears that cracking occurs only in the locally anisotropic case 
whereas in the perfect fluid case, once out of equilibrium, the configuration
 tends either to collapse or to expand as a whole. Furthermore it has been
shown in \cite{cn1}, that for a wide range of models, cracking appears only if, in the process of
perturbation leading to departure from equilibrium, the local anisotropy
is perturbed. This result suggests that  fluctuations of local anisotropy may
be the crucial factor in the occurrence of cracking. Thus, in the case of an initially locally isotropic configuration, the appearance of cracking shows that even small deviations from local isotropy
may lead to drastic changes in the evolution of the system as compared
with the purely locally isotropic case \cite{cn1}.

It is the purpose of this work to study in some detail the conditions under which, cracking appears in anisotropic polytropes. We shall consider the two possible cases of anisotropic polytropes mentioned above, and our models are the same as those considered in  \cite{2p}.
 
At this point, we have to make some comments about the perturbative
scheme, leading to departures from equilibrium. For the cracking to occur,
it is imperative that the perturbations introduced into the system
take it out of equilibrium. In other words we shall consider, exclusively,
perturbations under which the system is dynamically unstable. One way
to assure this, is to assume that the value of the ratio of specific heats
of the fluid is not equal to the critical value required for marginal (neutral)
dynamical stability. So, under perturbations, the configuration either
collapses or expands. Another completely equivalent way to achieve this
instability consists in assuming that, under perturbations of density and
local anisotropy, the radial pressure of the system maintains the same radial
dependence it had in equilibrium. It is obvious that in this case the hydrostatic equation will no longer be satisfied.

It should be clear that assuming this lack of response of the fluid,
i.e. the inability to adapt its radial pressure to the perturbed situation,
is equivalent to assuming that the pressure-density relation (the ratio of
specific heats) never reaches the value required for neutral equilibrium.

In this work, we shall consider two possible ways to perturb the energy density and the local anisotropy, namely:
\begin{itemize}
\item  We shall perturb the parameter $K$ and the parameter describing the anisotropy.

or
\item We shall perturb the parameter $n$ and the parameter describing the anisotropy.
\end{itemize}
As we shall see, cracking occurs for a wide range of values of the parameters.
\section{Relevant  Equations and Conventions}
\subsection{The field equations}
Even though polytropes are static configurations, we have to keep in mind that the appearance of cracking  occurs when the system is taken out of equilibrium. Accordingly, we shall consider spherically symmetric distributions of 
collapsing fluid, which we assume to be locally anisotropic,  bounded by a 
spherical surface $\Sigma$. The static limit is trivially recovered from the equations below.

\noindent
The line element is given in Schwarzschild-like coordinates by
 
\begin{equation}
ds^2=e^{\nu} dt^2 - e^{\lambda} dr^2 - 
r^2 \left( d\theta^2 + \sin^2\theta d\phi^2 \right),
\label{metric}
\end{equation}

\noindent
where $\nu(t,r)$ and $\lambda(t,r)$ are functions of their arguments. We 
number the coordinates: $x^0=t; \, x^1=r; \, x^2=\theta; \, x^3=\phi$.

\noindent 
which in our case read \cite{cn}:

\begin{equation}
-8\pi T^0_0=-\frac{1}{r^2}+e^{-\lambda} 
\left(\frac{1}{r^2}-\frac{\lambda'}{r} \right),
\label{feq00}
\end{equation}

\begin{equation}
-8\pi T^1_1=-\frac{1}{r^2}+e^{-\lambda}
\left(\frac{1}{r^2}+\frac{\nu'}{r}\right),
\label{feq11}
\end{equation}

\begin{eqnarray}
-8\pi T^2_2 & = & -  8\pi T^3_3 =  - \frac{e^{-\nu}}{4}\left[2\ddot\lambda+
\dot\lambda(\dot\lambda-\dot\nu)\right] \nonumber \\
& + & \frac{e^{-\lambda}}{4}
\left[2\nu''+\nu'^2 - 
\lambda'\nu' + 2\frac{\nu' - \lambda'}{r}\right],
\label{feq2233}
\end{eqnarray}

\begin{equation}
-8\pi T_{01}=-\frac{\dot\lambda}{r},
\label{feq01}
\end{equation}

\noindent
where dots and primes stand for partial differentiation with respect
to $t$ and $r$  
respectively.

\noindent
In order to give physical significance to the $T^{\mu}_{\nu}$ components 
we apply the Bondi approach \cite{Bo}.

\noindent
Thus, following Bondi, let us introduce purely locally Minkowski 
coordinates ($\tau, x, y, z$)

$$d\tau=e^{\nu/2}dt,\quad\,dx=e^{\lambda/2}dr,\quad\,
dy=rd\theta,\quad\, dz=r\sin\theta d\phi.$$

\noindent
Then, denoting the Minkowski components of the energy tensor by a bar, 
we have
\begin{eqnarray}
\bar T^0_0=T^0_0,\,\qquad\,
\bar T^1_1=T^1_1,\,\qquad\,\bar T^2_2=T^2_2,\,\nonumber \\
\bar T^3_3=T^3_3,\,\qquad\,\bar T_{01}=e^{-(\nu+\lambda)/2}T_{01}.
\label{bm}
\end{eqnarray}
\noindent
Next, we suppose that when viewed by an observer moving relative to these 
coordinates with proper velocity $\omega$ in the radial direction, the physical 
content  of space consists of an anisotropic fluid of energy density $\rho$, 
radial pressure $P_r$ and  tangential pressure $P_\bot$. Thus, when viewed by this moving observer the covariant tensor in 
Minkowski coordinates is

\[ \left(\begin{array}{cccc}
\rho     & 0&   0     &   0    \\
0 &  P_r    &   0     &   0    \\
0       &   0       & P_\bot  &   0    \\
0       &   0       &   0     &   P_\bot  
\end{array} \right). \]

\noindent
Then a Lorentz transformation readily shows that

\begin{equation}
T^0_0=\bar T^0_0= \frac{\rho + P_r \omega^2 }{1 - \omega^2}, 
\label{T00}
\end{equation}

\begin{equation}
T^1_1=\bar T^1_1=-\frac{ P_r + \rho \omega^2}{1 - \omega^2},
\label{T11}
\end{equation}

\begin{equation}
T^2_2=T^3_3=\bar T^2_2=\bar T^3_3=-P_\bot,
\label{T2233}
\end{equation}

\begin{equation}
T_{01}=e^{(\nu + \lambda)/2} \bar T_{01}=
-\frac{(\rho + P_r) \omega e^{(\nu + \lambda)/2}}{1 - \omega^2}.
\label{T01}
\end{equation}

\noindent
Note that the coordinate velocity in the ($t,r,\theta,\phi$) system, $dr/dt$, 
is related to $\omega$ by

\begin{equation}
\omega=\frac{dr}{dt}\,e^{(\lambda-\nu)/2}.
\label{omega}
\end{equation}

\noindent
Feeding back (\ref{T00}--\ref{T01}) into (\ref{feq00}--\ref{feq01}), we get the field equations in  the form

\begin{equation}
\frac{\rho + P_r \omega^2 }{1 - \omega^2} =-\frac{1}{r^2}+e^{-\lambda} 
\left(\frac{1}{r^2}-\frac{\lambda'}{r} \right),
\label{fieq00}
\end{equation}

\begin{equation}
\frac{ P_r + \rho \omega^2}{1 - \omega^2} =\frac{1}{r^2} - e^{-\lambda}
\left(\frac{1}{r^2}+\frac{\nu'}{r}\right),
\label{fieq11}
\end{equation}

\begin{eqnarray}
P_\bot = &  &\frac{e^{-\nu}}{4}\left[2\ddot\lambda+
\dot\lambda(\dot\lambda-\dot\nu)\right] \nonumber \\
& - & \frac{e^{-\lambda}}{4}
\left[2\nu''+\nu'^2 - 
\lambda'\nu' + 2\frac{\nu' - \lambda'}{r}\right],
\label{fieq2233}
\end{eqnarray}

\begin{equation}
\frac{(\rho + P_r) \omega e^{(\nu + \lambda)/2}}{1 - \omega^2} =\frac{\dot\lambda}{r}.
\label{fieq01}
\end{equation}

\noindent

At the outside of the fluid distribution, the spacetime is that of Schwarzschild, 
given by

\begin{equation}
ds^2= \left(1-\frac{2M}{{r}}\right) dt^2 -\frac{dr^2}{ \left(1-\frac{2M}{{r}}\right)}- 
r^2 \left(d\theta^2 + \sin^2\theta d\phi^2 \right).
\label{Vaidya}
\end{equation}

\noindent
In order to match smoothly the two metrics above on the boundary surface 
$r=r_\Sigma(t)$, we  require the continuity of the first and the second fundamental 
forms,  across that surface.

These last conditions imply
\begin{equation}
e^{\nu_\Sigma}=1-\frac{2M}{r_\Sigma},
\label{enusigma}
\end{equation}
\begin{equation}
e^{-\lambda_\Sigma}=1-\frac{2M}{r_\Sigma},
\label{elambdasigma}
\end{equation}
\begin{equation}
[P_r]_\Sigma=0,
\label{Psup}
\end{equation}
where, from now on, subscript $\Sigma$ indicates that the quantity is 
evaluated at the boundary surface $\Sigma$. Eqs. (\ref{enusigma}), (\ref{elambdasigma}) and (\ref{Psup}) are the necessary and 
sufficient conditions for a smooth matching of the two metrics (\ref{metric}) 
and (\ref{Vaidya}) on $\Sigma$.

The energy momentum tensor may be written as:
\begin{equation}
T_{\mu\nu} = \left(\rho+P_\bot\right)u_\mu u_\nu - P_\bot g_{\mu\nu} + 
\left(P_r-P_\bot\right)s_\mu s_\nu, 
\label{T-}
\end{equation}
where $u^\mu$ denotes the four velocity of the fluid and $s^\mu$ is a unit spacelike vector, radially directed. These vectors are defined by

\begin{equation}
u^\mu=\left[\frac{e^{-\nu/2}}{\left(1-\omega^2 \right)^{1/2}},\frac{\omega e^{-\lambda/2}}{\left(1-\omega^2 \right)^{1/2}},0,0\right],
\label{umu}
\end{equation}
\begin{equation}
s^\mu=\left[\frac{\omega \, e^{-\nu/2}}{\left(1-\omega^2\right)^{1/2}},\,
\frac{e^{-\lambda/2}}{\left(1-\omega^2\right)^{1/2}},\,0,\,0\right].
\label{smu}
\end{equation}

\noindent
Next, it will be useful to calculate the radial component of the 
``conservation law''

\begin{equation}
T^\mu_{\nu;\mu}=0,
\label{dTmn}
\end{equation}

\noindent
which in the static case becomes

\begin{equation}
R\equiv P'_r+\frac{\nu'}{2}\left(\rho+P_r\right)-
\frac{2\left(P_\bot-P_r\right)}{r}=0,
\label{Prp}
\end{equation}

\noindent
representing the generalization of the Tolman-Oppenheimer-Volkof equation 
for anisotropic fluids \cite{14}.
Alternatively, using 
\begin{equation}
\frac{\;\nu'}{2} = \frac{m + 4 \pi P_r r^3}{r \left(r - 2m\right)},
\label{nuprii}
\end{equation}
which follows from (\ref{fieq00}) and (\ref{fieq11}) in the static case, we may write
\begin{equation}
P'_r=-\frac{(m + 4 \pi P_r r^3)}{r \left(r - 2m\right)}\left(\rho+P_r\right)+\frac{2\left(P_\bot-P_r\right)}{r},\label{ntov}
\end{equation}
 where   the mass function $m(r)$, as usually, is  defined by 
\begin{equation}
e^{-\lambda}=1-2m/r, \qquad m(r) = 4 \pi \int^r_0{\rho r^2 dr}.\label{mass}
\end{equation}

Polytropes  are static fluid configurations, which satisfy either of  the equation (\ref{p1})  or (\ref{p2}). The full set of equations describing the structure of these self-gravitating objects, in both cases, were derived and discussed in \cite{2p}.

All the models   have to satisfy physical requirements such as:
\begin{equation}
\rho >0, \qquad  \frac{P_r}{\rho}\leq 1, \qquad  \frac{P_\perp }{\rho} \leq 1.
\label{conditions}
\end{equation}

In what follows, we shall very briefly review the main equations corresponding to each case.

\subsection{Case I}
Assuming Eq. (\ref{p1}), let us introduce  the following variables
\begin{equation}
\alpha=P_{rc}/\rho_{c},\quad r=\xi/A,  \quad A^2=4 \pi \rho_{c}/\alpha (n+1)\label{alfa},\end{equation}

\begin{equation}\Psi_{0}^n=\rho_{0}/\rho_{0c},\quad v(\xi)=m(r) A^3/(4 \pi\rho_{c}),\label{psi}\end{equation}
where subscript $c$ indicates that the quantity is evaluated at the center. At the boundary surface $r=r_\Sigma$ ($\xi=\xi_\Sigma$) we have $\Psi_0(\xi_\Sigma)=0$. 

Then, the generalized Tolman-Opphenheimer-Volkoff equation becomes
\begin{widetext}
\begin{eqnarray}
\xi^2 \frac{d\Psi_{0}}{d\xi}\left[\frac{1-2(n+1)\alpha v/\xi}{(1-n\alpha)+(n+1)\alpha \Psi_{0}}\right]+v+\alpha\xi^3 \Psi_{0}^{n+1}-\frac{2\Delta \Psi_0^{-n}\xi}{P_{rc}(n+1)} \left[\frac{1-2\alpha (n+1)v/\xi}{(1-n\alpha)+(n+1)\alpha \Psi_0}\right]=0,\label{TOV1anis_WB}
\end{eqnarray}
\end{widetext}
where $\Delta=P_\bot-P_r$.

On the other hand we obtain from the mass function definition (\ref{mass})  and Eq. (\ref{fieq00}),
\begin{equation}
m'=4 \pi r^2 \rho,\label{mprima}\end{equation}
or
\begin{equation}
\frac{dv}{d\xi}=\xi^2 \Psi_{0}^n (1-n\alpha+n\alpha\Psi_{0}).\label{veprima}\end{equation}

In this case, conditions (\ref{conditions}) read
\begin{eqnarray}
n\alpha<1,\qquad  \frac{\alpha \Psi_0}{1-n\alpha+n\alpha \Psi_0}\leq 1, \nonumber \\ \frac{3v/\xi^3+\alpha \Psi_0^{n+1}}{\Psi_0^n(1-n\alpha+n\alpha\Psi_o)}-1\leq1.
\label{conditionsII}
\end{eqnarray}

\subsection{Case II}

In this case the assumed equation of state is (\ref{p2}), then, introducing
\begin{equation}
\Psi^n=\rho/\rho_{c},\label{psi2}\end{equation}
the generalized Tolman-Opphenheimer-Volkoff equation becomes

\begin{widetext}

\begin{eqnarray}
\xi^2 \frac{d\Psi}{d\xi}\left[\frac{1-2(n+1)\alpha v/\xi}{1+\alpha \Psi}\right]+v+\alpha\xi^3 \Psi^{n+1}-\frac{2\Delta \Psi^{-n}\xi}{P_{rc}(n+1)} \left[\frac{1-2\alpha(n+1)v/\xi}{1+\alpha \Psi}\right]=0,\label{TOV2anis_WB}
\end{eqnarray}
\end{widetext}
and from {Eq.} (\ref{mprima})
\begin{equation}\frac{dv}{d\xi}=\xi^2 \Psi^n.\label{veprima2}\end{equation}
In this case, conditions (\ref{conditions}) read:
\begin{equation}
\rho>0, \qquad \alpha \Psi \leq 1, \qquad \frac{3v}{\xi^3 \Psi^n}+\alpha \Psi-1 \leq1.
\label{conditionsIII}
\end{equation}

Equations (\ref{TOV1anis_WB}), (\ref{veprima}) or (\ref{TOV2anis_WB}), (\ref{veprima2}), form a system of two first order  ordinary differential equations for the three unknown functions: $\Psi (\Psi_0), v, \Delta$, depending on a duplet of parameters $n, \alpha$. Thus, it is obvious that in order to proceed further with the modeling of a compact object, we need to provide additional information. Such  information, of course, depends on the specific physical problem under consideration. Here,  we shall further assume the equation of state used in \cite{anis4, 2p}.

\section{Perturbing the anisotropic polytrope}
Let us now consider an anisotropic polytrope satisfying the equation of hydrostatic equilibrium (\ref{Prp}). Furthermore, our distribution satisfies the equation of state proposed in \cite{anis4}, i.e.

\begin{equation}
\Delta =C(\rho +P_r)\left[\frac{m+4\pi P_r r^3}{r(r-2m)}\right]r,
\label{2}
\end{equation}
where $C$ is a constant, producing

\begin{equation}
R=\frac{dP_r}{dr}+h\left[ \frac{m+4\pi r^3 P_r}{r(r-2m)}\right]  (\rho +P_r)=0,
\label{3}
\end{equation}
with

\begin{equation}
h\equiv 1-2C.
\label{h}
\end{equation}

In order to observe an eventual cracking, we have to take the system out of equilibrium. For doing so we shall  perturb the energy density and the anisotropy, through  the parameters $K$ and $h$, or, alternatively, through  the parameters $n$ and $h$. Below we describe this procedure for the cases I and II.

\subsection{Perturbation in the case I}
In this case the following equations apply

\begin{equation}
P_r=K\rho^{1+1/n}_0 \quad \mbox{and} \quad \rho=\rho_0+nK\rho_0^{1+1/n}.
\label{4}
\end{equation}

Then, perturbing the energy density and the local anisotropy via $K$ and  $h$, 

\begin{eqnarray}
K \longrightarrow \tilde{K} &=& K+\delta K ,\label{5} \\
h \longrightarrow \tilde{h} &=& h+\delta h, \label{6} 
\end{eqnarray} 
it follows that
\begin{equation}
\tilde{P}_r=\tilde{K}\rho_0^{1+1/n}=\beta P_r,
\label{8}
\end{equation}

\begin{equation}
\tilde{\rho}=\rho_0 + n\tilde{K}\rho_0^{1+1/n}=\rho_0 +n\beta P_r,
\end{equation}

with

\begin{equation}
\beta = \frac{\tilde{K}}{K},
\label{9}
\end{equation}
 where the tilde denotes the perturbed quantity.

Introducing the dimensionless variable

\begin{equation}
\hat{\tilde{R}}=\frac{A}{4\pi \rho_c^2}\tilde{R},
\label{12}
\end{equation}
we obtain from (\ref{4})-(\ref{9})

\begin{equation}
\hat{\tilde{R}}=\beta \Psi_0^n \frac{d\Psi_0}{d\xi}+\frac{\tilde{h} \Psi^n_0}{\xi^2}\left[(1-n\alpha)+\alpha \beta (n+1)\Psi_0 \right]\left[\frac{\tilde{\upsilon}+\xi^3 \beta \alpha \Psi_0^{n+1}}{1-2\alpha (n+1)\tilde{\upsilon}/\xi}\right].
\label{13}
\end{equation}

From the above, it follows that up to first order, we may write
\begin{widetext}
\begin{equation}
\delta \hat{R} = \hat{\tilde{R}}(\xi, 1+\delta \beta,h+\delta h, \upsilon + \delta \upsilon) 
= \hat{R}(\xi ,1,h,\upsilon)+\left. \left(\frac{\partial \hat{\tilde{R}}}{\partial \beta }\right)\right|_{\begin{array}{c}
\beta =1 \\
\tilde{h}= h \\
\tilde{\upsilon}=\upsilon
\end{array}}. \delta \beta + \left. \left(\frac{\partial \hat{\tilde{R}}}{\partial \tilde{\upsilon}}\right)\right|_{\begin{array}{c}
\beta =1 \\
\tilde{h}= h \\
\tilde{\upsilon}=\upsilon
\end{array}}. \delta \upsilon + \left. \left(\frac{\partial \hat{\tilde{R}}}{\partial \tilde{h}}\right)\right|_{\begin{array}{c}
\beta =1 \\
\tilde{h}= h \\
\tilde{\upsilon}=\upsilon
\end{array}}. \delta h.
\label{sn1}
\end{equation}
\end{widetext}
where

\begin{equation}
\hat{\tilde{R}}(\xi, 1,h, \upsilon)=0,
\label{sn2}
\end{equation}
as it must be, since the system is at equilibrium in the unperturbed state.

Thus, we may write
\begin{widetext}
\begin{eqnarray}
\delta \hat{R}=\left. \left(\frac{\partial \hat{\tilde{R}}}{\partial \beta }\right)\right|_{\begin{array}{c}
\beta =1 \\
\tilde{h}= h \\
\tilde{\upsilon}=\upsilon
\end{array}}. \delta \beta + \left. \left(\frac{\partial \hat{\tilde{R}}}{\partial \tilde{\upsilon}}\right)\right|_{\begin{array}{c}
\beta =1 \\
\tilde{h}= h \\
\tilde{\upsilon}=\upsilon
\end{array}}. \delta \upsilon + \left. \left(\frac{\partial \hat{\tilde{R}}}{\partial \tilde{h}}\right)\right|_{\begin{array}{c}
\beta =1 \\
\tilde{h}= h \\
\tilde{\upsilon}=\upsilon
\end{array}}. \delta h.
\label{14}
\end{eqnarray}
\end{widetext}

Then, using  (\ref{13}) we obtain
\begin{widetext}
\begin{eqnarray}
\left. \left(\frac{\partial \hat{\tilde{R}}}{\partial \beta }\right)\right|_{\begin{array}{c}
\beta =1 \\
\tilde{h}= h \\
\tilde{\upsilon}=\upsilon
\end{array}} &=& \Psi_0^n \frac{d\Psi_0}{d\xi}+h\frac{\Psi_0^{n+1}}{\xi^2}\alpha \left[\frac{(n+1)(\upsilon +2\xi^3 \alpha \Psi_0^{n+1})+\xi^3 \Psi_0^{n}(1-n\alpha)}{1-2\alpha (n+1)\upsilon /\xi} \right], \\
\left. \left(\frac{\partial \hat{\tilde{R}}}{\partial \tilde{\upsilon}}\right)\right|_{\begin{array}{c}
\beta =1 \\
\tilde{h}= h \\
\tilde{\upsilon}=\upsilon
\end{array}} &=& h\frac{\Psi_0^n}{\xi^2}\left[\frac{(1-n\alpha)+\alpha (n+1)\Psi_0}{(1-2\alpha(n+1)\upsilon /\xi)^2}\right][1+2\alpha^2 (n+1) \xi^2 \Psi_0^{n+1}], \\
\left. \left(\frac{\partial \hat{\tilde{R}}}{\partial \tilde{h}}\right)\right|_{\begin{array}{c}
\beta =1 \\
\tilde{h}= h \\
\tilde{\upsilon}=\upsilon
\end{array}} &=& \frac{\Psi_0^n}{\xi^2}\left[(1-n\alpha)+\alpha (n+1)\Psi_0 \right]\left[\frac{\upsilon +\xi^3 \alpha \Psi_0^{n+1}}{1-2\alpha (n+1)\upsilon /\xi} \right].
\end{eqnarray}
\end{widetext}
Furthermore, we have
\begin{equation}
\tilde{\upsilon}=\frac{1}{\rho_c}\int_{0}^{\xi}\bar{\xi}^2 \Psi^n_0(\rho_{0c}+n\beta P_{rc}\Psi_0)d\bar{\xi},
\end{equation}
and
\begin{equation}
\delta \upsilon =\left. \left(\frac{\partial \tilde{\upsilon}}{\partial \beta}\right) \right|_{\beta=1}.\delta \beta,
\end{equation}
producing
\begin{equation}
\delta \upsilon = n\alpha \left(\int_{0}^{\xi}\bar{\xi}^2 \Psi_0^{n+1}d\bar{\xi}\right)\delta \beta = n\alpha F(\xi)\delta \beta,
\end{equation}
with
\begin{equation}
F(\xi)\equiv \int_{0}^{\xi}\bar{\xi}^2 \Psi_0^{n+1}d\bar{\xi}.
\label{e7n}
\end{equation}
From the equations above, we may write
\begin{widetext}
\begin{equation*}
\delta \hat{R}  =  \Psi_0^n \left\{\frac{d\Psi_0}{d\xi} +\frac{h \alpha}{\xi^2(1-2\alpha (n+1)\upsilon /\xi)}\left[(n+1)(\upsilon +2\xi^3\alpha \Psi_0^{n+1})\Psi_0 + \xi^3\Psi_0^{n+1}(1-n\alpha) \right. \right.
\end{equation*}
\begin{equation*}
+\left. \left. \left[\frac{(1-n\alpha)+\alpha (n+1)\Psi_0}{1-2\alpha (n+1)\upsilon /\xi}\right]\left[1+2\alpha^2 (n+1)\xi^2 \Psi_0^{n+1}\right]nF(\xi)\right] \right\}\delta \beta
\end{equation*}
\begin{equation}
+ \frac{\Psi_0^n}{\xi^2} \left[(1-n\alpha)+\alpha (n+1)\Psi_0\right]\left[\frac{\upsilon+\xi^3 \alpha \Psi_0^{n+1}}{1-2\alpha (n+1)\upsilon /\xi}\right]\delta h.
\label{e8n}
\end{equation}
\end{widetext}
It would be more   convenient to use the variable $x$, defined by
\begin{equation}
x=\frac{\xi}{\bar{A}}, \quad \bar{A}=r_{\Sigma}A= \xi_{\Sigma},
\label{nxn}
\end{equation}
in terms of which, (\ref{e8n}) becomes
\begin{widetext}
\begin{eqnarray}
\delta \hat{{R}} & = & \frac{\Psi_0^n}{\bar{A}} \left\{\frac{d\Psi_0}{dx} + \frac{h \alpha}{x(\bar{A}x - 2\alpha (n+1)\upsilon)}\left[(n+1)(\upsilon + 2\bar{A}^3 x^3 \alpha \Psi_0^{n+1})\Psi_0 \bar{A}^3x^3 \Psi_0^{n+1} (1-n\alpha) \right. \right. \nonumber \\[1em] 
& + & \left. \left. \bar{A}x \left[\frac{1-n\alpha +\alpha (n+1)\Psi_0}{\bar{A}x -2\alpha (n+1)\upsilon}\right](1+2\alpha^2 (n+1)\bar{A}^2x^2 \Psi_0^{n+1})nF_1 \right] \right\} \delta \beta \nonumber \\[1em]
& + & \frac{\Psi_0^n}{\bar{A}x}\left(1-n\alpha + \alpha (n+1)\Psi_0 \right)\left[\frac{\upsilon +\bar{A}^3x^3\Psi_0^{n+1}}{\bar{A}x - 2\alpha (n+1)\upsilon}\right]\delta h.
\label{e8nb}
\end{eqnarray}
\end{widetext}
Now,  for the cracking to occur, once the system has been perturbed, there must be a change of sign in $\delta \hat{R}$. More specifically, it should be positive in the inner regions, and negative at the outer ones (in the inverse case an overturning is produced). This last condition implies that $\delta \hat{R}=0$ for some value of  $\xi$ in the interval $[0,\xi_\Sigma]$, implying in turn:

\begin{equation}
\delta h=-\Gamma \delta \beta,
\end{equation}
where

\begin{eqnarray}
\Gamma= \left.\left[ \frac{\frac{\partial \tilde{R}}{\partial \beta}+\frac{\partial \tilde{R}}{\partial \tilde{\upsilon}}F(\xi)}{\frac{\partial \tilde{R}}{\partial \tilde{h}}} \right]\right|_{\begin{array}{c}
\beta=1 \\
\tilde{\upsilon}=\upsilon \\
\tilde{h}=h
\end{array}}.
\end{eqnarray}

Another perturbative scheme may be proposed, based on the perturbation of the  energy density and the anisotropy through the parameters $n$ and  $h$. In  this case, following the same line of arguments as in the previous scheme, we have

\begin{eqnarray}
P_r &=& k\rho_{0}^{1+1/n} = k\rho_{0c}^{1+1/n}\Psi_0^{n+1} \\
\rho &=& \rho_0 +nP_r = \rho_{0c}\Psi_0^{n} + nk\rho_{0c}^{1+1/n}\Psi_0^{n+1} 
\end{eqnarray}

\begin{eqnarray}
n \longrightarrow \tilde{n} &=& n+\delta n \\
h \longrightarrow \tilde{h} &=& h+\delta h. 
\end{eqnarray}
\\
Then, assuming that the radial pressure remains unchanged under perturbation (what  ensures the departure from equilibrium)

\begin{equation}
\tilde{P_r}=P_r,
\end{equation}

we have

\begin{equation}
\tilde{P_r}=P_r=k\rho_{0c}^{1+1/n}\Psi^{n+1},
\end{equation} 

\begin{equation}
\tilde{\rho}=\rho_{0c}\Psi^{\tilde{n}}+\tilde{n}P_r.
\end{equation}

Then, replacing into the hydrostatic equlibrium equation, it follows that
\begin{widetext}
\begin{equation}
\hat{\tilde{R}}=\Psi_0^n\frac{d\Psi_0}{d\xi}+\frac{\tilde{h}}{\xi^2}[(1-n\alpha)\Psi_0^{\tilde{n}}+(\tilde{n}+1)\alpha \Psi_0^{n+1}]\left[\frac{\tilde{\upsilon}+\xi^3\alpha \Psi_0^{n+1}}{(1-2\alpha (n+1)\tilde{\upsilon}/\xi)}\right].
\end{equation}
\end{widetext}
Proceeding exactly as in the previous scheme, we obtain
\begin{widetext}
\begin{eqnarray}
\delta \hat{{R}}&=&\frac{h \Psi_0^n}{\xi^2\left(1-2\alpha (n+1)\upsilon /\xi\right)}\left[(\upsilon +\xi^3\alpha \Psi_0^{n+1})((1-n\alpha)\ln (\Psi_0)+\alpha \Psi_0)  +\left[\frac{1-n\alpha +(1+n)\alpha \Psi_0}{1-2\alpha (n+1)\upsilon /\xi}\right](1+2\alpha^2\xi^2(n+1)\Psi_0^{n+1})F_{1n}\right] \delta n \nonumber \\
&+&\frac{\Psi_0^n}{\xi^2}[1-n\alpha +(n+1)\alpha \Psi_0]\left[\frac{\upsilon +\xi^3\alpha \Psi_0^{n+1}}{1-2\alpha (n+1)\upsilon /\xi}\right]\delta h,
\label{e6n}
\end{eqnarray}
\end{widetext}
or,
\begin{widetext}
\begin{eqnarray}
\delta \hat{{R}} & = & \frac{h\Psi_0^n}{\bar{A}x(\tilde{A}x - 2\alpha (n+1)\upsilon)} \left[ (\upsilon +\bar{A}^3x^3\alpha \Psi_0^{n+1})((1-n\alpha)\ln(\Psi_0)+\alpha \Psi_0) \right. \nonumber\\[1em]
& + & \left.   \bar{A}x \left[\frac{1-n\alpha +\alpha (n+1)\Psi_0}{\bar{A}x -2\alpha (n+1)\upsilon}\right](1+2\alpha^2 (n+1)\bar{A}^2x^2 \Psi_0^{n+1})F_{1n}\right]\delta n \nonumber\\[1em]
& + & \frac{\Psi_0^n}{\bar{A}x}\left(1-n\alpha + \alpha (n+1)\Psi_0 \right)\left[\frac{\upsilon +\bar{A}^3x^3\alpha \Psi_0^{n+1}}{\bar{A}x - 2\alpha (n+1)\upsilon}\right]\delta h,
\label{e6nb}
\end{eqnarray}
\end{widetext}
with

\begin{equation}
F_{1n}=\int_0^\xi {\bar\xi}^2 \Psi_0^n [(1-n\alpha)\ln(\Psi_0)+\alpha \Psi_0]d\bar \xi.
\label{e5n}
\end{equation}

Again, for the cracking to occur we must have $\delta \hat{{R}}=0$  for some value of  $\xi$  in the interval $[0,\xi_\Sigma]$, implying :

\begin{equation}
\delta h=-\Gamma \delta n,
\end{equation}

with

\begin{eqnarray}
\Gamma= \left.\left[ \frac{\frac{\partial \tilde{R}}{\partial \tilde{n}}+\frac{\partial \tilde{R}}{\partial \tilde{\upsilon}}F_{1n}}{\frac{\partial \tilde{R}}{\partial \tilde{h}}} \right]\right|_{\begin{array}{c}
\tilde{n}=n \\
\tilde{\upsilon}=\upsilon \\
\tilde{h}=h.
\end{array}}
\end{eqnarray}

\subsection{Perturbation in the case II}

For this kind of polytropes, we have
\begin{equation}
 P_r=K\rho^{1+1/n},\quad \rho=\frac{\rho_0}{(1+K\rho^{1/n})^{n}}.
\end{equation}

Then,  we shall write for the radial pressure

\begin{equation}
P_r=K\rho_c^{1+1/n}G, 
\end{equation}
with

\begin{equation}
G \equiv \left(\frac{\rho}{\rho_c}\right)^{1+1/n},
\end{equation}
thereby ensuring that the radial dependence of $P_r$ remains unchanged after perturbation.

We shall now proceed to perturb our polytrope, first through the parameters  $K$ and  $h$. Then, following the same line of arguments as for the case I, we have

\begin{equation}
\tilde{P_r}=\tilde{K}\rho_c^{1+1/n}G=\beta P_r,
\end{equation} 

\begin{equation}
\tilde{\rho}=\frac{\rho_0}{(1+\tilde{K}\rho_0^{1/n})^n}=\frac{\rho_0}{(1+\beta K\rho_0^{1/n})^n},
\end{equation}
or, using

\begin{eqnarray}
\tilde{\rho} &=& \rho +\delta \rho =\rho +\left(\frac{\partial \tilde{\rho}}{\partial \beta}\right)_{\beta=1}\delta \beta \\
\beta &=& 1+\delta \beta,
\end{eqnarray}
we obtain

\begin{equation}
\tilde{\rho}=\rho +nP_r(\beta -1)
\end{equation}
Feeding back the above expressions into (\ref{3}) we may write

\begin{equation}
\tilde{\hat{R}}=\beta \Psi^n \frac{d\Psi}{d\xi}+\frac{\Psi^n \tilde{h}}{\xi^2}(1+\alpha \Psi[(n+1)\beta -n])\frac{\tilde{\upsilon}+\xi^3 \alpha \beta \Psi^{n+1}}{(1-2(n+1)\alpha \tilde{\upsilon})/\xi}.
\end{equation}

Carrying out a procedure similar to the one described in the previous subsection we obtain
\begin{widetext}
\begin{eqnarray}
\delta \hat{{R}} & = & \frac{\Psi^n}{\bar{A}} \left\{\frac{d\Psi }{dx} + \frac{h \alpha}{x(\bar{A}x - 2\alpha (n+1)\upsilon)}\left[ \Psi (\upsilon (n+1)+ \bar{A}^3x^3\Psi^{n}(1+\alpha (n+2)\Psi)) \right. \right. \nonumber\\[1em] 
& + & \left. \left. \bar{A}x\left(\frac{1+ \alpha \Psi}{\bar{A}x - 2\alpha (n+1)\upsilon} \right)(1+2(n+1)\alpha^2\bar{A}^2x^2\Psi^{n+1})nF_2 \right] \right\}\delta \beta 
 +  \frac{\Psi^n}{\bar{A}x}(1+\alpha \Psi)\left[\frac{\upsilon + \bar{A}^3x^3\alpha \Psi^{n+1}}{\bar{A}x - 2\alpha (n+1)\upsilon)}\right]\delta h.
\label{grafI}
\end{eqnarray}
\end{widetext}
where (\ref{nxn}) has been used, and

\begin{equation}
F_2=\int_0^{\xi}{\bar \xi^2} \Psi^{n+1}d{\bar\xi}.
\label{e2n}
\end{equation}

If the perturbation is introduced via the parameter  $n$, then the result is: 
\begin{widetext}
\begin{eqnarray}
\delta \hat{{R}} & = & \frac{h \Psi^n}{x(\bar{A}x - 2\alpha (n+1)\upsilon)}\left[(\upsilon + \bar{A}^3x^3\alpha \Psi^{n+1})\ln(\Psi)
 +  \bar{A}x\left(\frac{1+ \alpha \Psi}{\bar{A}x - 2\alpha (n+1)\upsilon} \right)(1+2(n+1)\alpha^2\bar{A}^2x^2\Psi^{n+1})F_{2n} \right] \delta n \nonumber \\[1em]
& + & \frac{\Psi^n}{\bar{A}x}(1+\alpha \Psi)\left[\frac{\upsilon + \bar{A}^3x^3\alpha \Psi^{n+1}}{\bar{A}x - 2\alpha (n+1)\upsilon)}\right]\delta h,
\label{e3n}
\end{eqnarray}
\end{widetext}

with

\begin{equation}
F_{2n}=\int^\xi_0 \ln (\Psi)\Psi^n d\bar\xi.
\label{e1n}
\end{equation}
\begin{figure}
\includegraphics[width=3.in,height=3.in,angle=0]{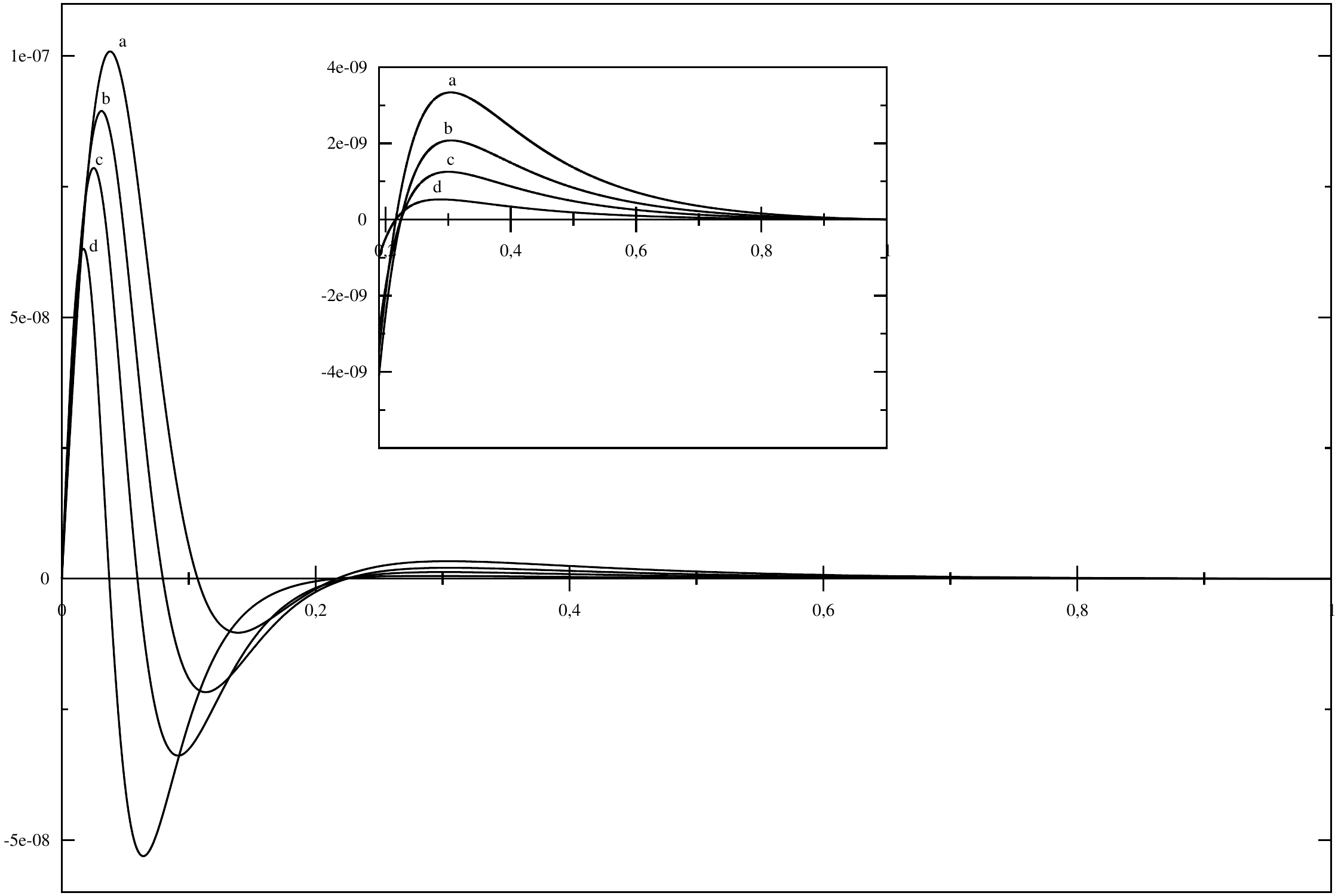} 
\caption{$\delta \hat{R}$ as a function of $x$ for $n=1$, $h=1.5$ and $\Gamma=1.6$. Curves a, b, c and d correspond to $\alpha=.83, \ .85, \ .87 \ \mbox{and} \ .90$, respectively.}
\label{fig:wd}
\end{figure}
\begin{figure}
\includegraphics[width=2.8in,height=2.5in,angle=0]{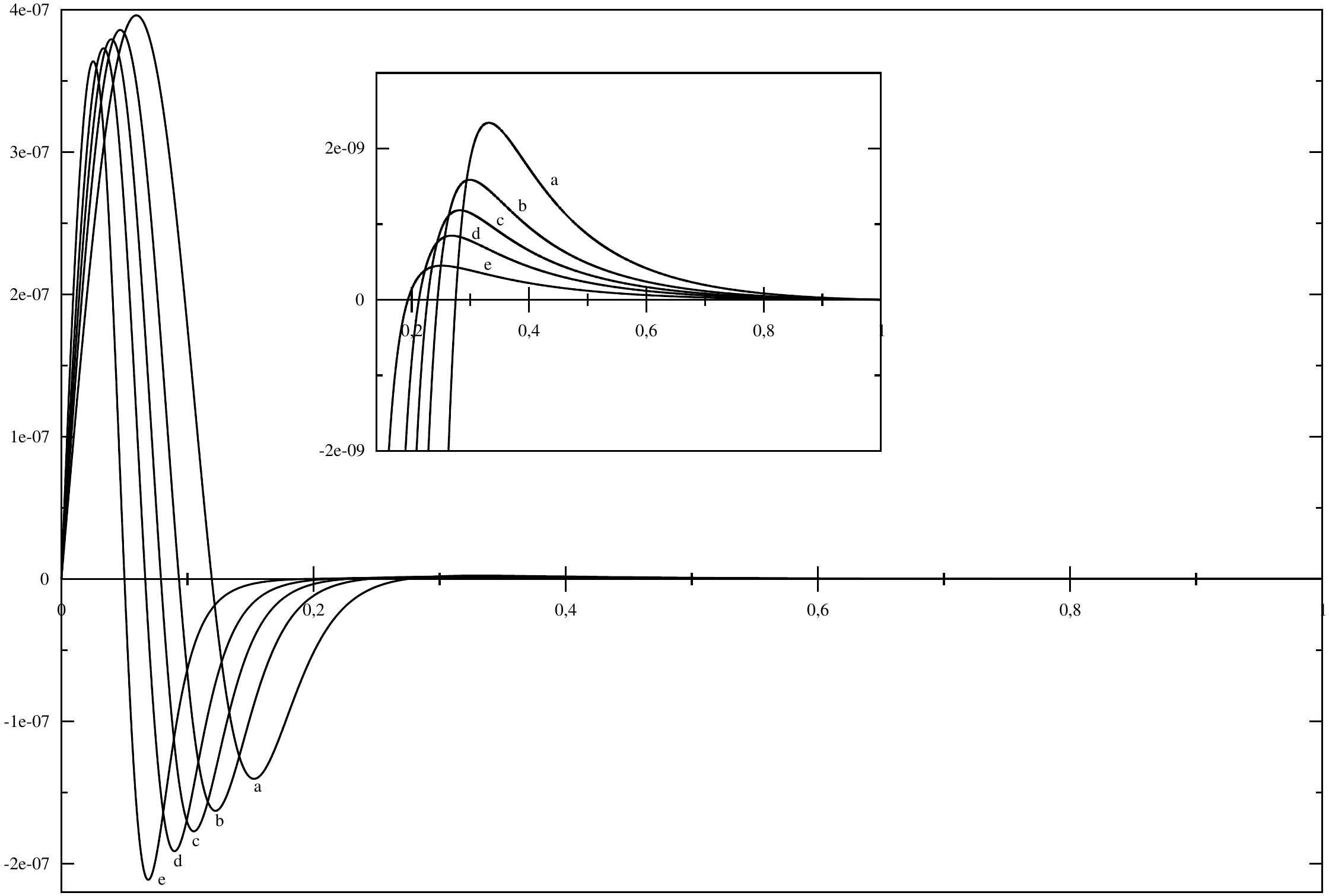} 
\caption{$\delta \hat{R}$ as a function of $x$ for $n=1$, $h=.5$ and $\Gamma=1$.  Curves a, b, c, d and e correspond to $\alpha=.80, \ .83, \ .85, \ .87 \ \mbox{and} \ .90$, respectively.}
\label{fig:mass}
\end{figure}
\begin{figure}
\includegraphics[width=3.5in,height=4.in,angle=0]{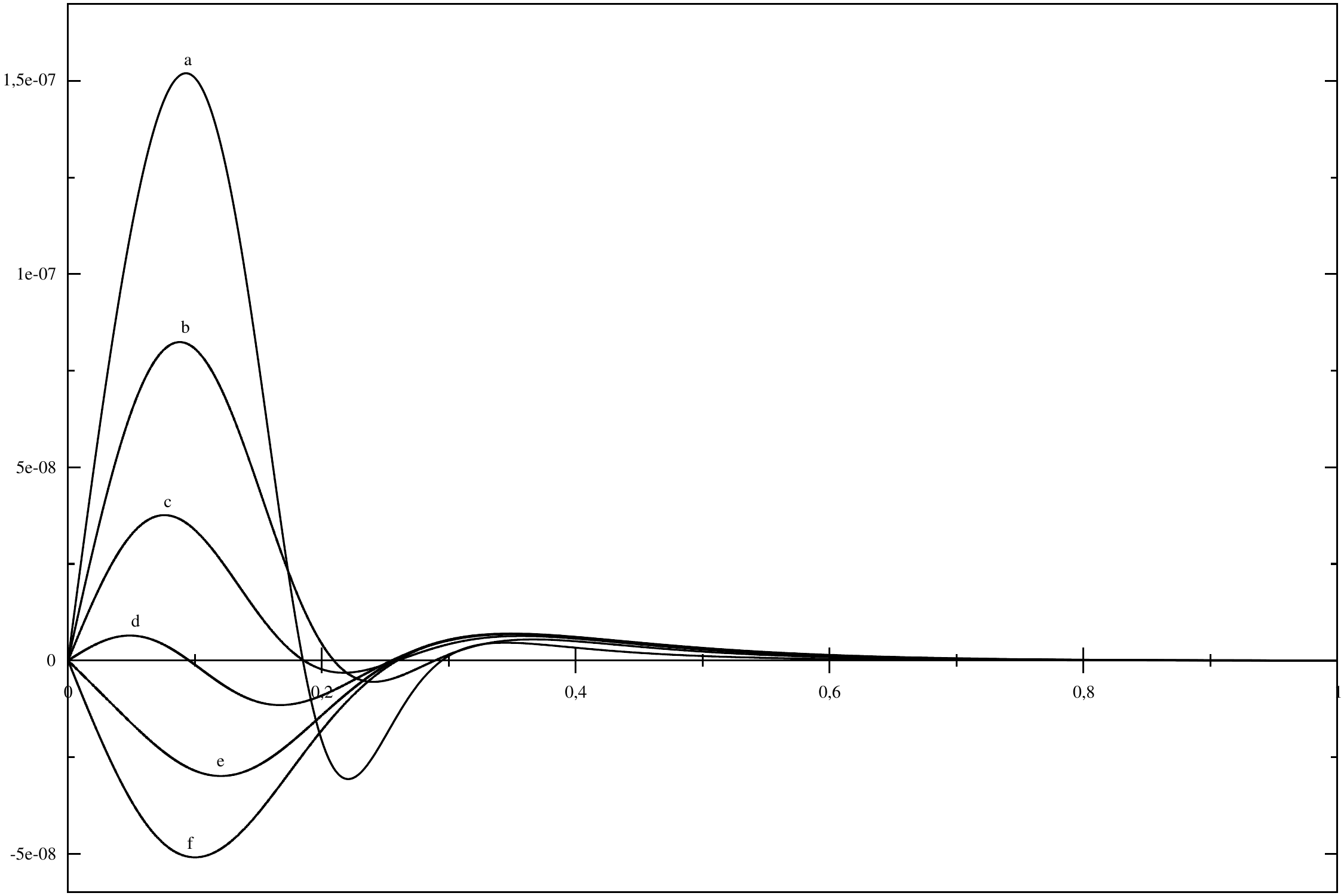} 
\caption{$\delta \hat{R}$ as a function of $x$ for $n=1.5$, $\alpha=.390$, $\Gamma=.6$ and different values of $h$. Curves a, b, c, d, e and f correspond to $h=.5, \ .7, \ .9, \ 1.1, \ 1.3, \ \mbox{and} \ 1.5$, respectively.}
\label{fig:isothermic}
\end{figure}
\begin{figure}
\includegraphics[width=3.5in,height=4.in,angle=0]{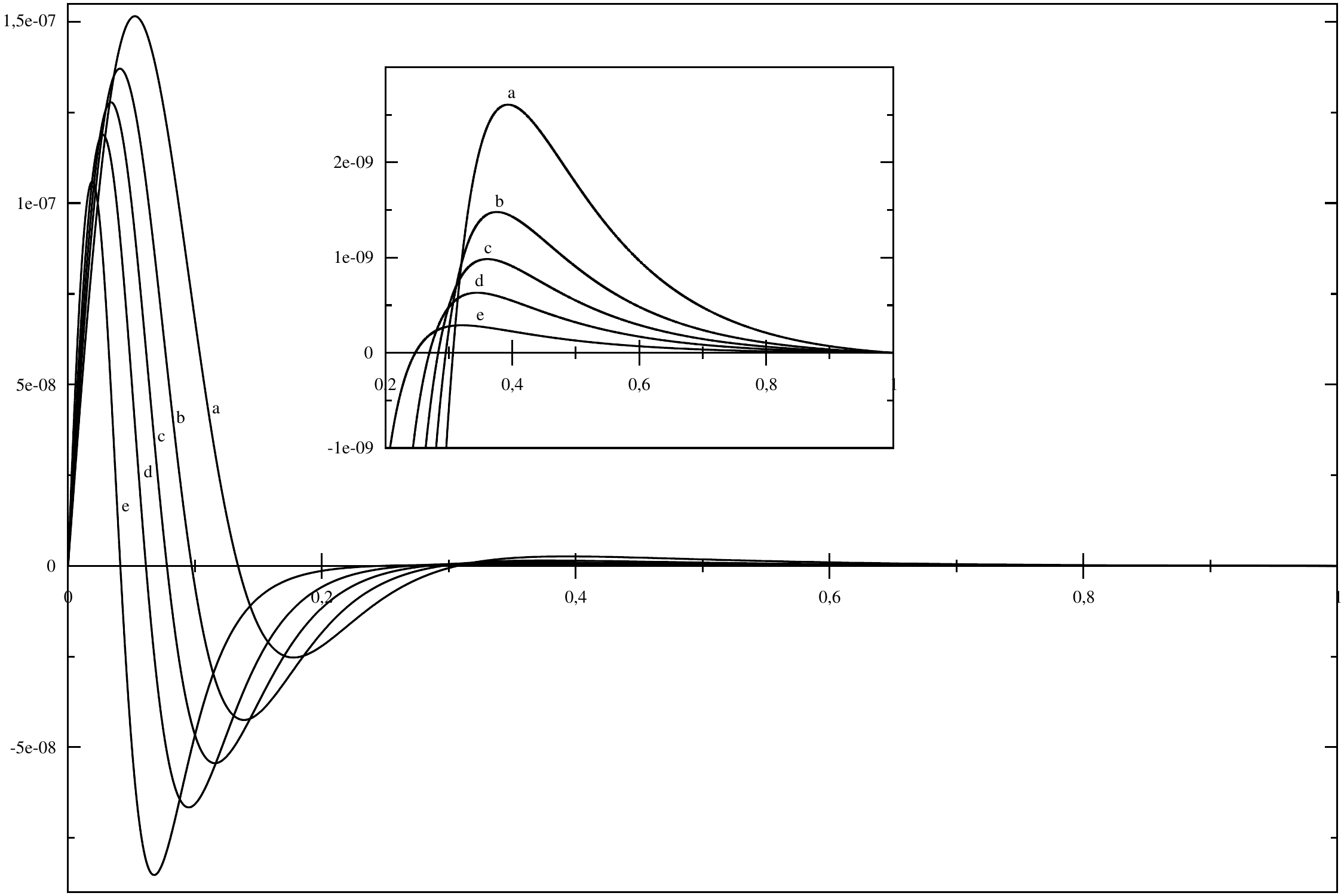} 
\caption{$\delta \hat{R}$ as a function of $x$ for $n=1$, $h=1$  and $\Gamma=1.2$. Curves a, b, c, d and e correspond to $\alpha=.80, \ .83, \ .85, \ .87 \ \mbox{and} \ .90$, respectively.}
\label{fig:isothermic}
\end{figure}
\begin{figure}
\includegraphics[width=3.5in,height=4.in,angle=0]{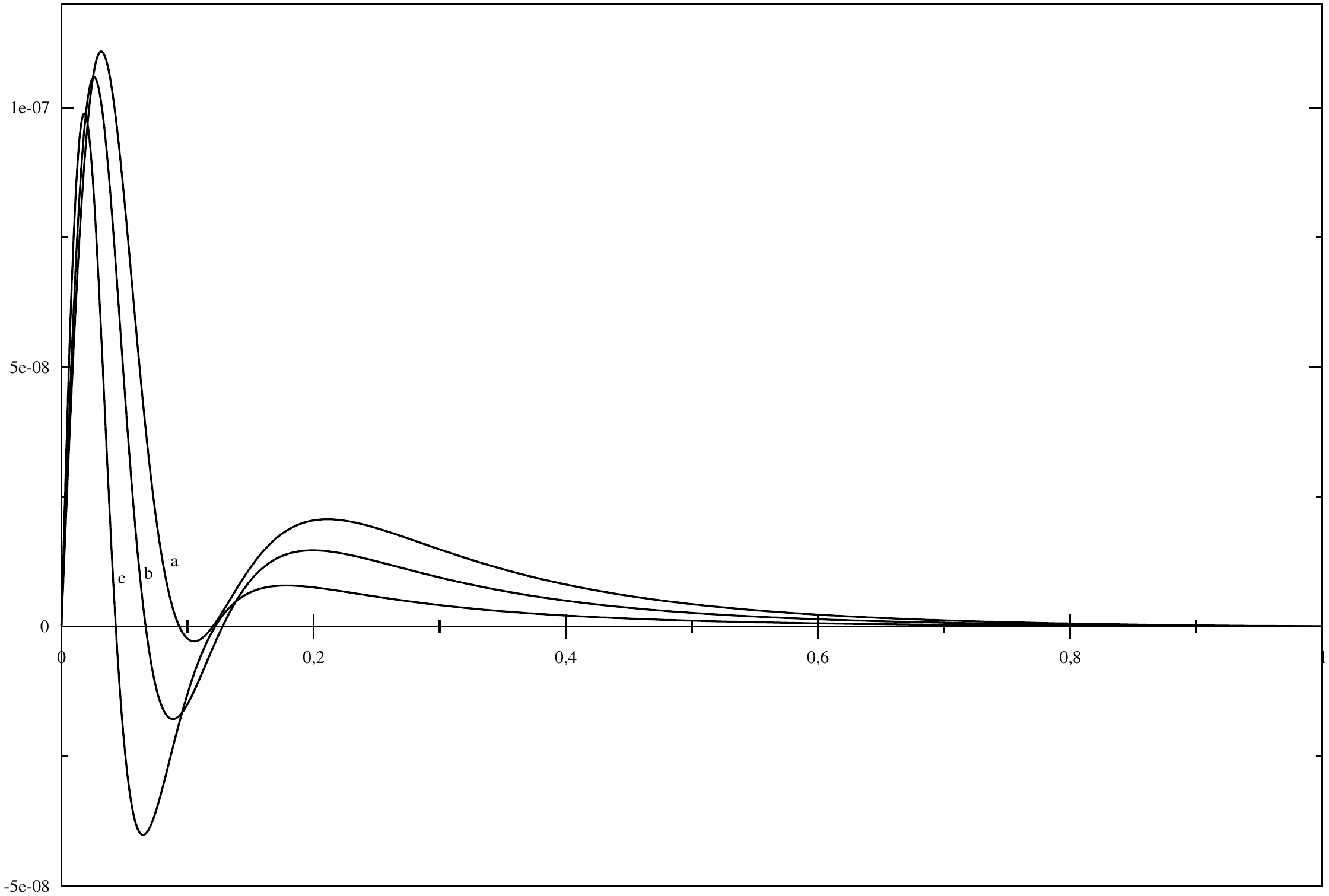} 
\caption{$\delta \hat{R}$ as a function of $x$ for $n=1$, $h=1.5$  and $\Gamma=1.4$. Curves a, b and c correspond to $\alpha=.85, \ .87 \ \mbox{and} \ .90$, respectively.}
\label{fig:isothermic}
\end{figure}
\begin{figure}
\includegraphics[width=3.5in,height=4.in,angle=0]{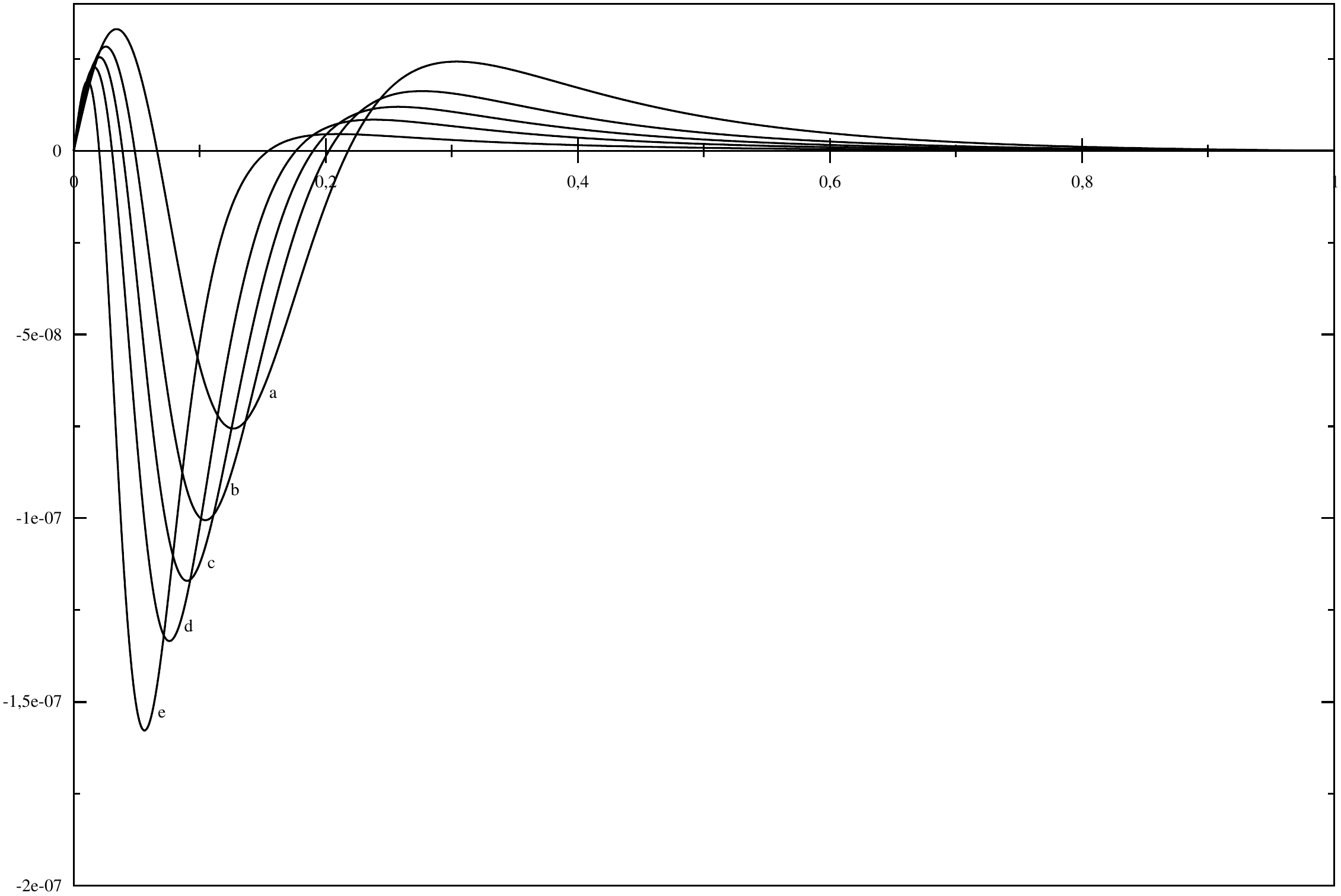} 
\caption{$\delta \hat{R}$ as a function of $x$ for $n=1$, $h=1$  and $\Gamma=.8$. Curves a, b, c, d and e correspond to $\alpha=.80, \ .83, \ .85, \ .87 \ \mbox{and} \ .90$, respectively.}
\label{fig:isothermic}
\end{figure}
\begin{figure}
\includegraphics[width=3.5in,height=4.in,angle=0]{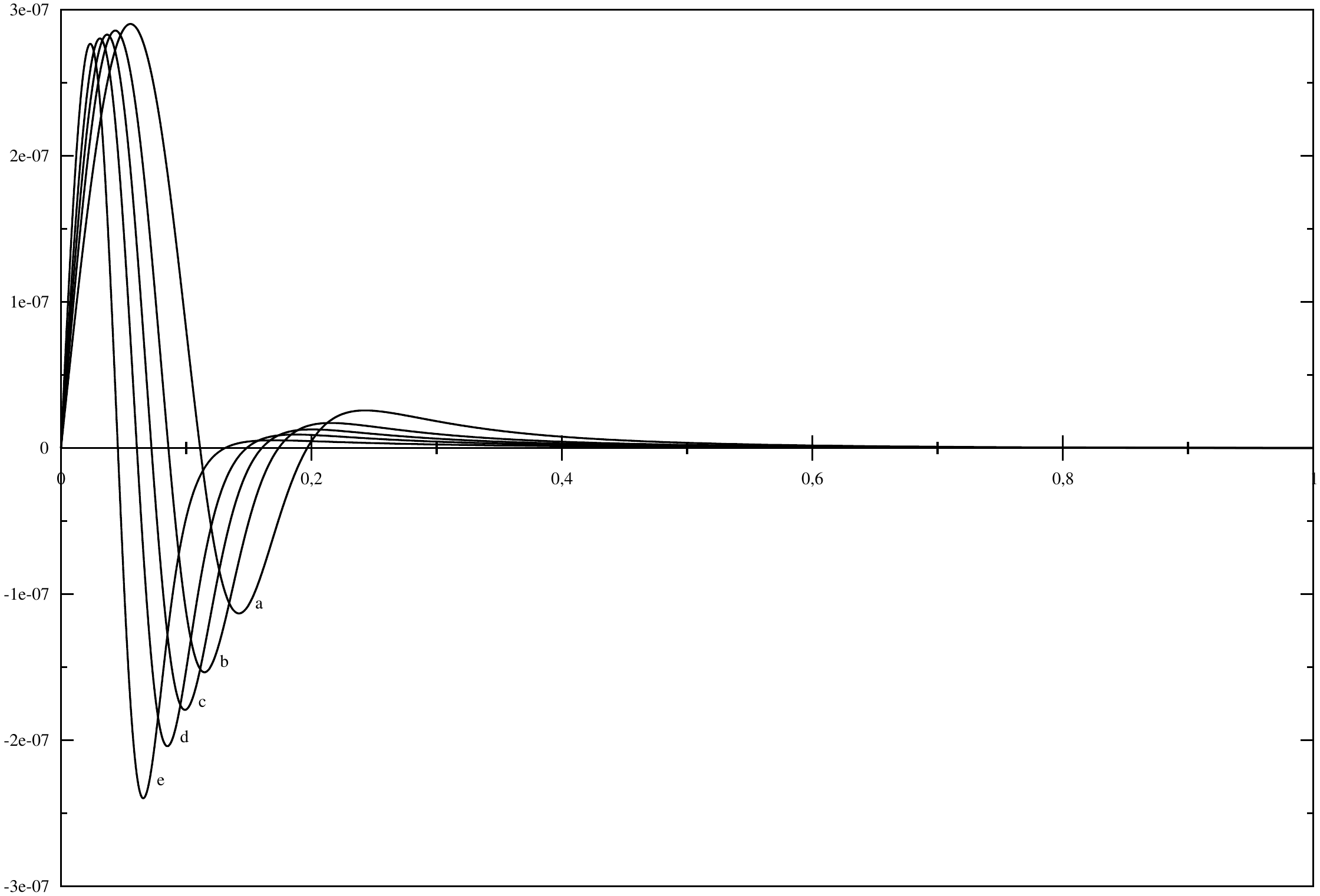} 
\caption{$\delta \hat{R}$ as a function of $x$ for $n=1$, $h=.5$ and $\Gamma=.8$. Curves a, b, c, d and e correspond to $\alpha=.80, \ .83, \ .85, \ .87 \ \mbox{and} \ .90$, respectively.}
\label{fig:isothermic}
\end{figure}
\section{Results}
We shall now proceed to apply the formalism sketched in the previous section, to detect and analyze the occurrence of cracking in the family of polytropes defined by the equation of state (\ref{2}). We shall consider both types of polytropes (I and II), and both schemes of perturbation (through $K$ and $h$, and through $n$ and $h$).

To do so, using a four order Runge--Kutta method,  we shall integrate (\ref{TOV1anis_WB}), (\ref{veprima}), and  (\ref{TOV2anis_WB}), (\ref{veprima2}), for any  duplet of parameters $n, \alpha$ and $n, h$ , for which there exists a boundary surface. Such duplets have been determined in \cite{2p} (see figures (3, 4) for the case I, and (8, 9) for the case II, in that reference). Also, the absolute numerical values of the perturbations of the parameters were taken of the order of $10^{-6}$, the results are not qualitatively sensitive, to  changes in  some orders of magnitude, below or above, that figure. Finally, the integrals (\ref{e7n}, \ref{e6n}, \ref{e2n}, \ref{e1n}) were numerically evaluated by means of the trapezoid rule (see for example \cite{trap}).

Next, the obtained $\Psi, \Psi_0$ (which of course coincide with the results in \cite{2p}) are used to evaluate (\ref{e8nb}) and (\ref{e6nb}) for the case I, and (\ref{grafI}) and (\ref{e3n}) for the case II. 
We have considered the whole range of values of the parameters leading to bounded configurations, whose physical variables satisfy the physical requirements (\ref{conditions}).

We shall now analyze the obtained results for each case.
\begin{figure}
\includegraphics[width=3.in,height=4.in,angle=0]{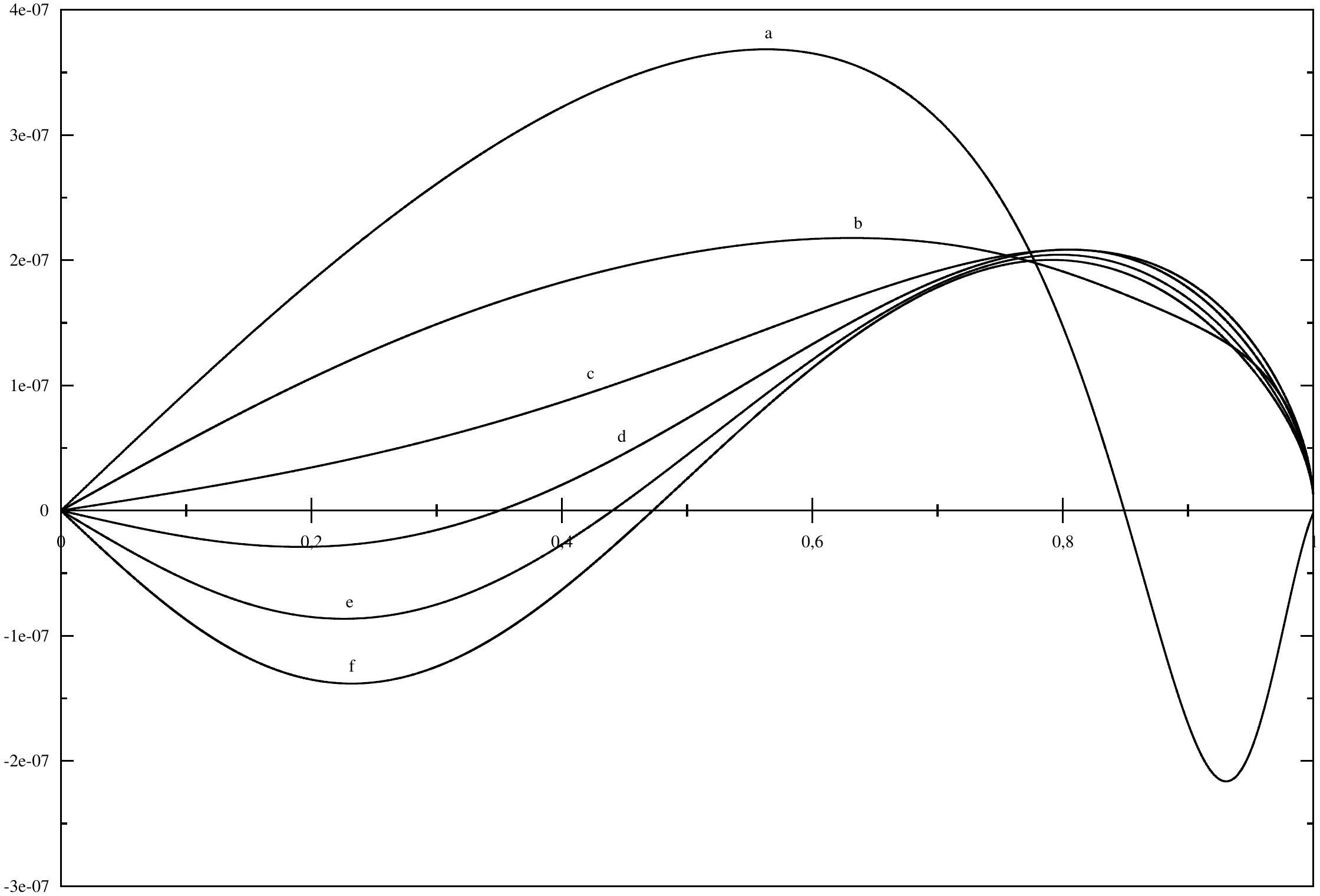} 
\caption{$\delta \hat{R}$ as a function of $x$ for $n=.5$, $\alpha=.98$, $\Gamma=.6$ and different values of $h$. Curves a, b, c, d, e and f correspond to $h=.5, \ .7, \ .9, \ 1.1, \ 1.3, \ \mbox{and} \ 1.5$, respectively.}
\label{fig:isothermic}
\end{figure}
\begin{figure}
\includegraphics[width=3.in,height=4.in,angle=0]{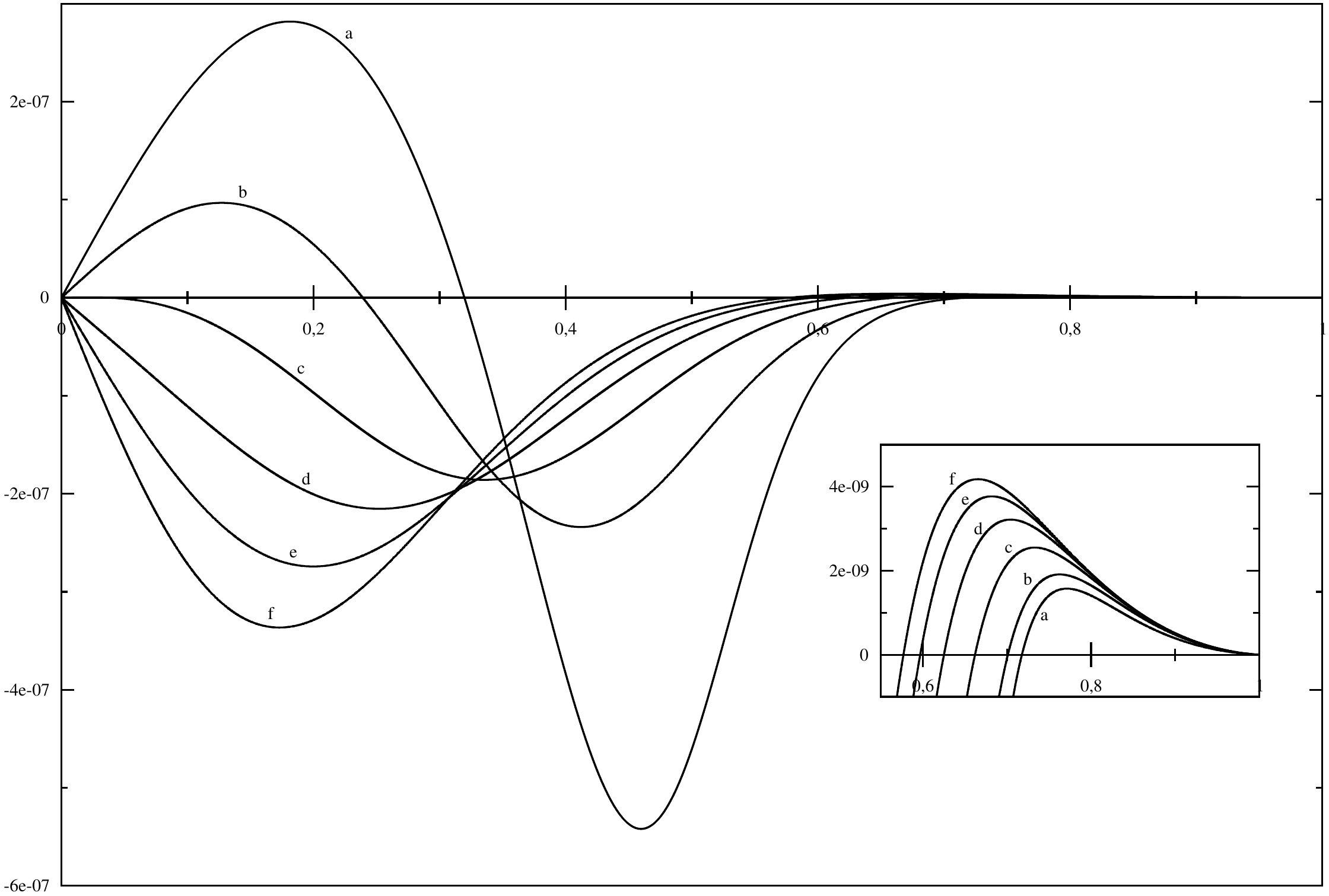} 
\caption{$\delta \hat{R}$ as a function of $x$ for $n=1.5$, $\alpha=.84$, $\Gamma=1.1$ and different values of $h$. Curves a, b, c, d, e and f correspond to $h=.5, \ .7, \ .9, \ 1.1, \ 1.3, \ \mbox{and} \ 1.5$, respectively.}
\label{fig:isothermic}
\end{figure}
\begin{figure}
\includegraphics[width=3.in,height=4.in,angle=0]{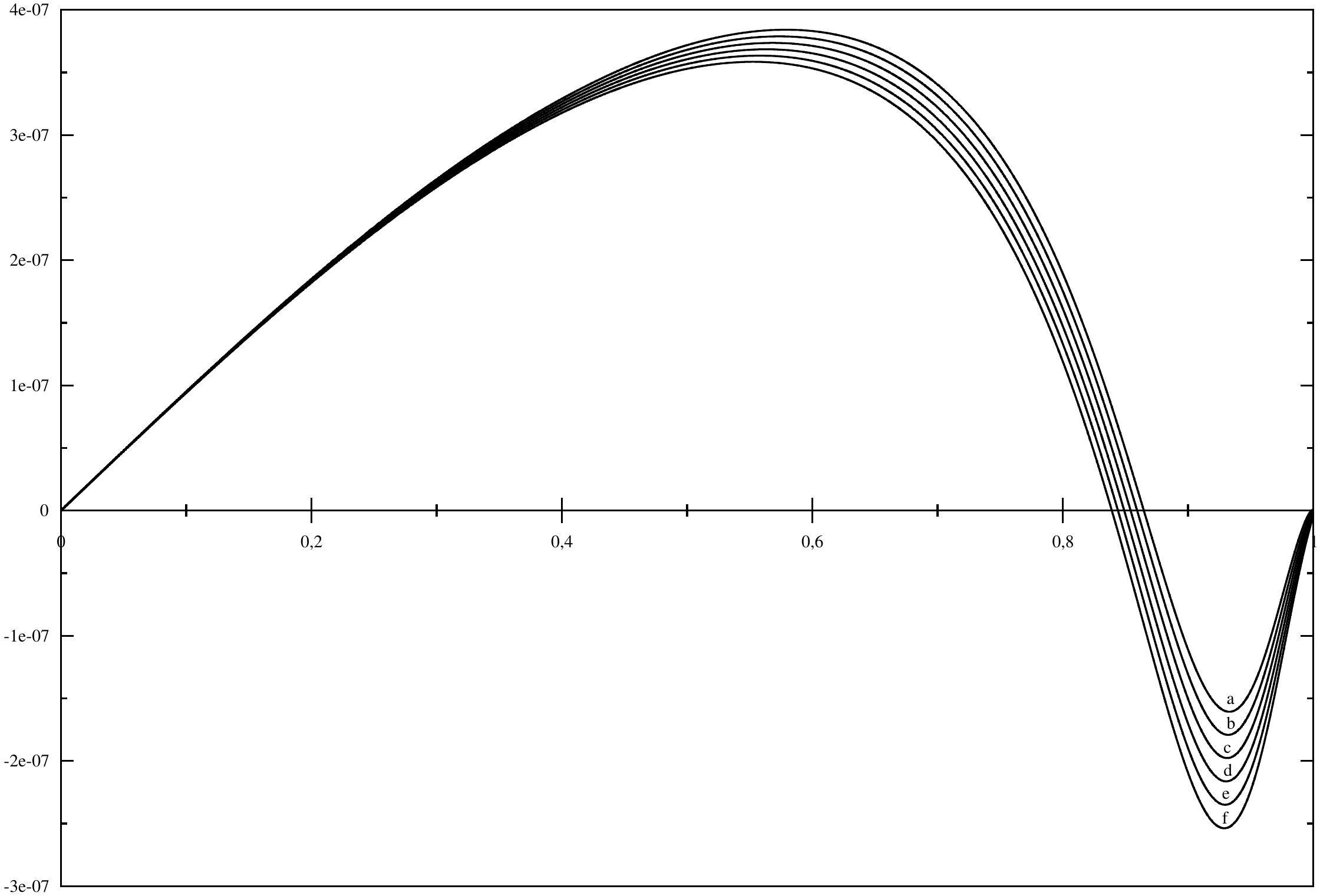} 
\caption{$\delta \hat{R}$ as a function of $x$ for $n=.5$, $h=.5$  and $\Gamma=.6$. Curves a, b, c, d, e and f correspond to $\alpha=.95, \ .96, \ .97, \ .98, \ .99 \ \mbox{and} \ 1.00$, respectively. }
\label{fig:isothermic}
\end{figure}
\begin{figure}
\includegraphics[width=3.in,height=4.in,angle=0]{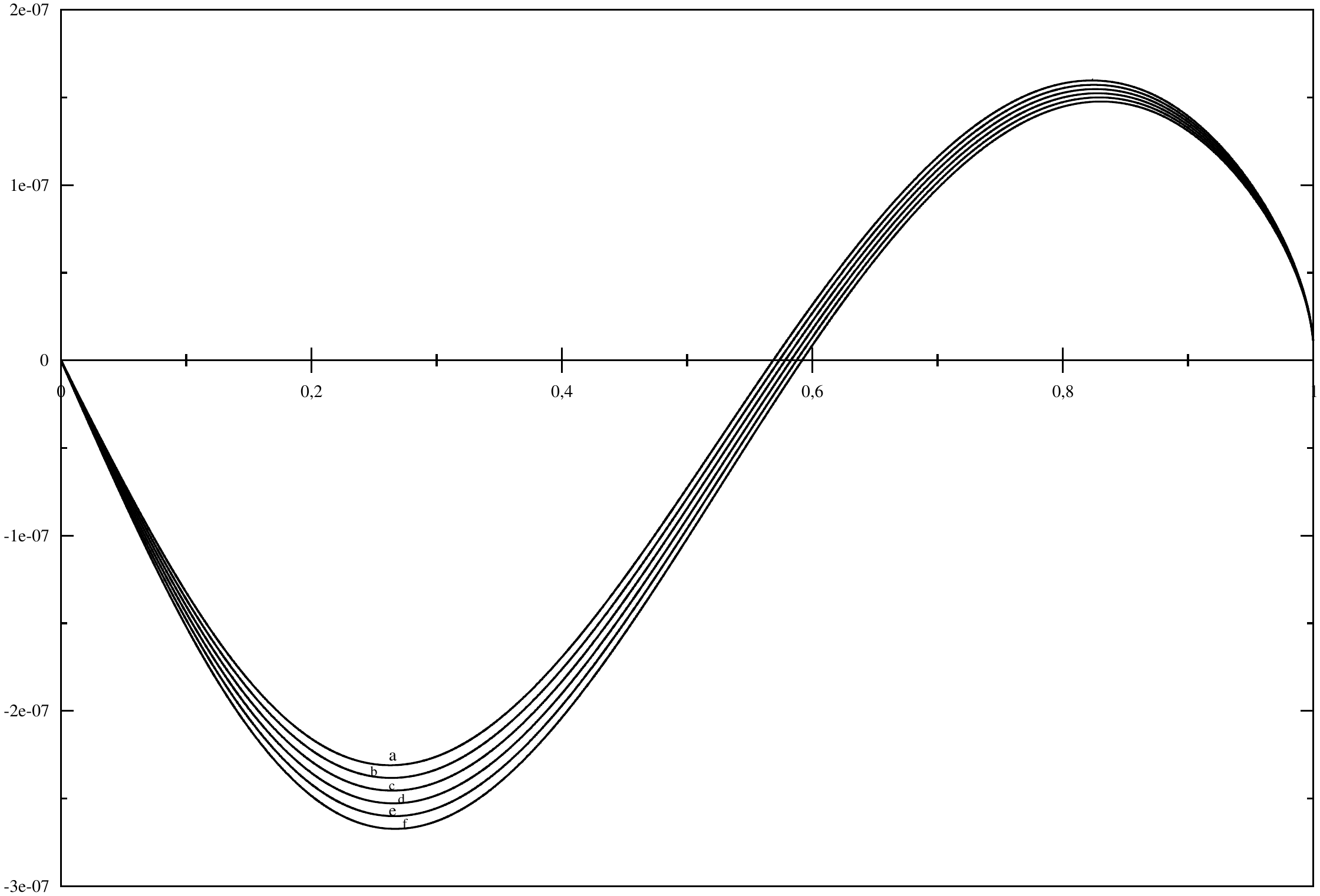} 
\caption{$\delta \hat{R}$ as a function of $x$ for $n=.5$, $h=1.5$  and $\Gamma=.4$. Curves a, b, c, d, e and f correspond to $\alpha=.95, \ .96, \ .97, \ .98, \ .99 \ \mbox{and} \ 1.00$, respectively.}
\label{fig:isothermic}
\end{figure}
\begin{figure}
\includegraphics[width=3.5in,height=4.in,angle=0]{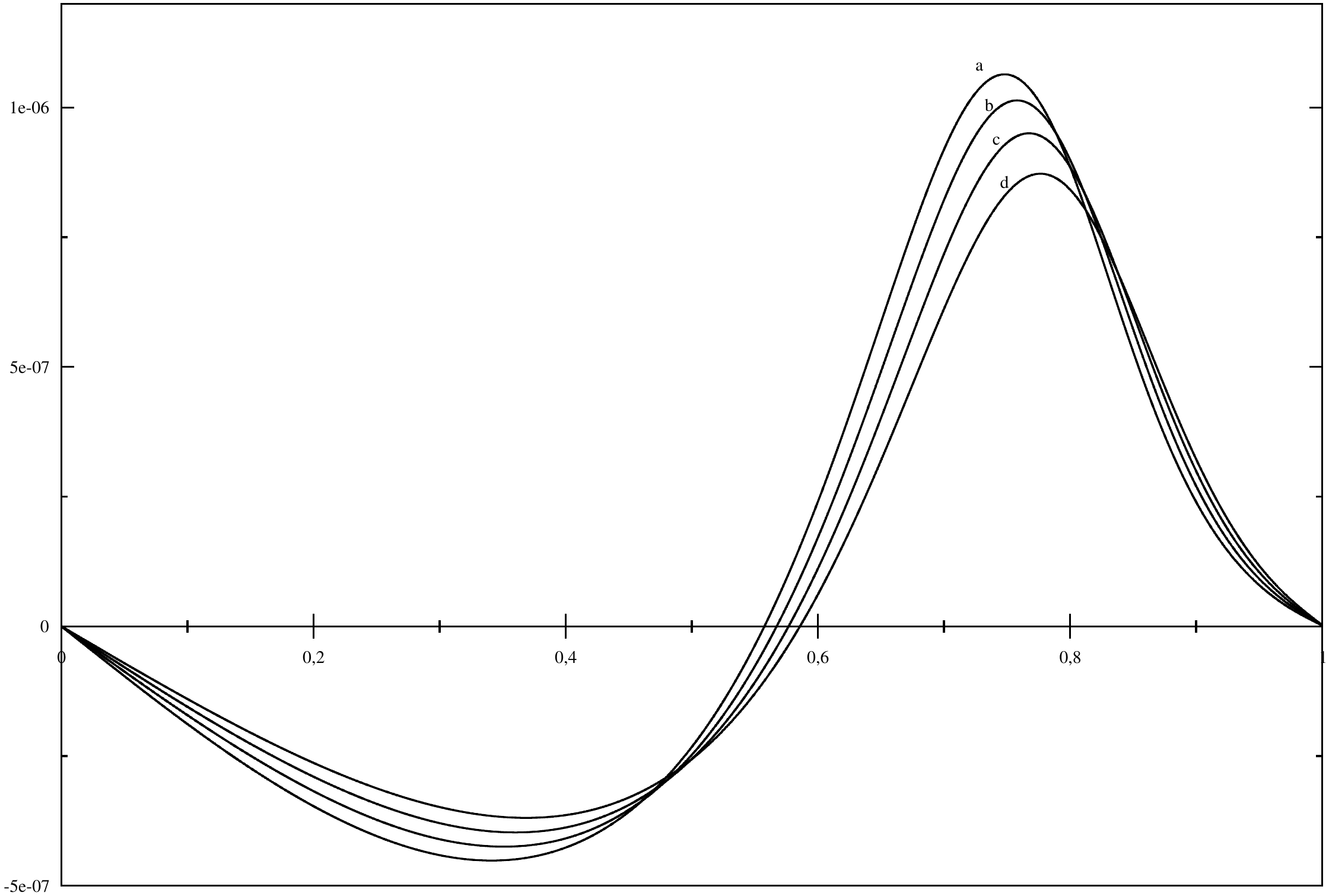} 
\caption{$\delta \hat{R}$ as a function of $x$ for $n=1$, $h=.5$  and $\Gamma=-0.5$. Curves a, b, c, and d correspond to $\alpha=1.0, \ .9, \ .8, \ \mbox{and} \ .7$, respectively.}
\label{fig:isothermic}
\end{figure}
\begin{figure}
\includegraphics[width=3.5in,height=4.in,angle=0]{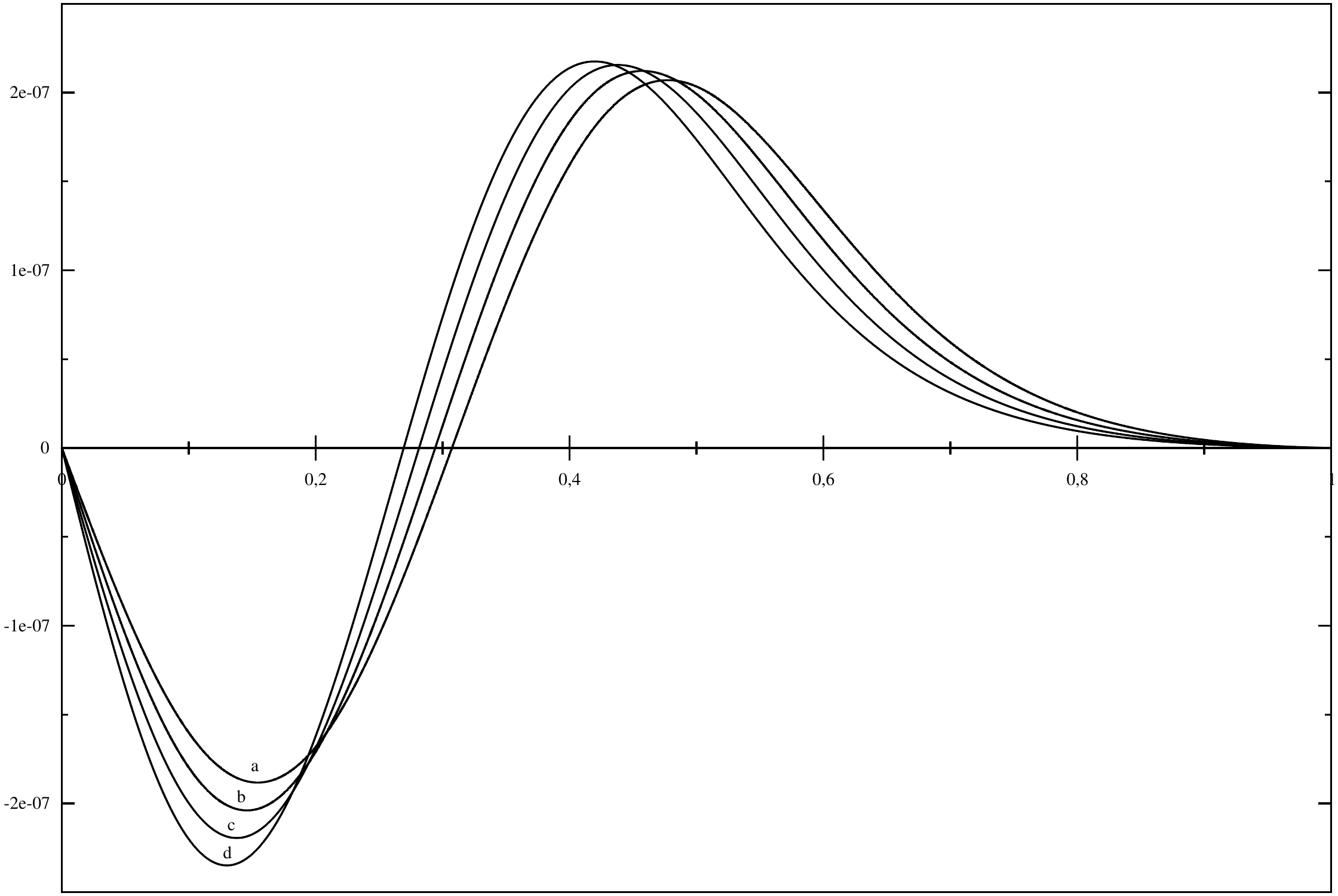} 
\caption{$\delta \hat{R}$ as a function of $x$ for $n=1.5$, $h=1$  and $\Gamma=-0.5$. Curves a, b, c, and d correspond to $\alpha=.7, \ .8, \ .9, \ \mbox{and} \ 1.0$, respectively.}
\label{fig:isothermic}
\end{figure}
\begin{figure}
\includegraphics[width=3.5in,height=4.in,angle=0]{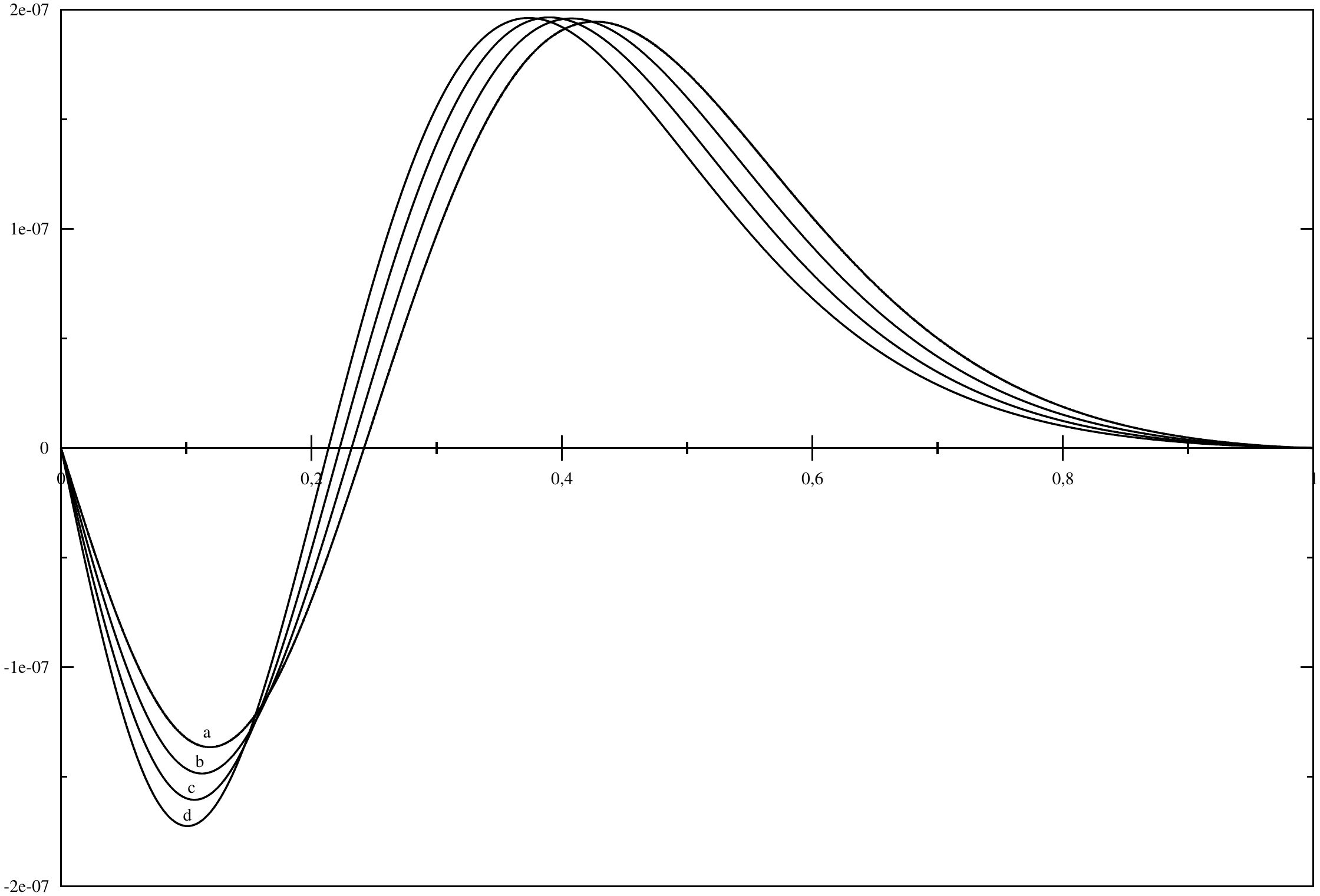} 
\caption{$\delta R$ as a function of $x$ for $n=1.5$, $h=1.5$  and $\Gamma=-0.5$.Curves a, b, c, and d correspond to $\alpha=.7, \ .8, \ .9, \ \mbox{and} \ 1.0$, respectively.}
\label{fig:isothermic}
\end{figure}
\subsection{Cracking for polytropes of type I}
Figures (1--7) summarize the main results obtained for the polytropes of type I. The first four figures (figs. (1--4)) correspond to the case where the perturbation is carried on through the parameter $K$, whereas  the remaining three figures of this case (figs.(5--7)) describe the behaviour of the system which has been perturbed through the parameter $n$.

In figure 1 we  observe the cracking fo $h=1.5$ and  the different values of $\alpha$ indicated in the figure legend. It appears that the strongest, and deepest, crackings are associated with the largest values of $\alpha$. We also observe that in the outer regions, the models  exhibit overturnings. In these latter cases, smallest values of  $\alpha$ are associated with strongest overturnings.

Figure 2 depicts a similar situation, but for $h=.5$.  However in this case crackings are stronger  and overturnings are weaker.

The situation described in figures 1 and 2 is representative for a wide range of parameteres (for which there exist bounded configurations satisfying the required physical conditions).

Figure 3 exhibits further  the close  relationship, existing between the occurrence of cracking and the type of anisotropy.  Indeed, the smallest values of $h$ correspond to the  deepest and strongest  crackings, and to the  weakest overturnings. Furthermore, for the largest values of $h$ (curves e and f) there are not crackings at all, only overturnings. 

Finally figure 4 exhibits the occurrence of cracking for configurations that are locally isotropic before perturbation, thereby illustrating the fact  that even slight deviations from local isotropy,  may be sufficient to produce a cracking.

Figures 5--7, which describe the results of perturbation through the parameter $n$, lead to conclusions qualitatively similar to the ones mentioned above. The strongest and deepest crackings, and the weakest overturnings,  are associated with largest values of $\alpha$

\subsection{Cracking for polytropes of type II}
The results of this case are depicted by figures (8--14). The first four of these, correspond to the perturbative scheme through the parameter $K$ whereas the last three correspond to perturbations of $n$.

Figure 8 describes the situation, for different values of $h$. For the choice of parameters in this figure, we observe that cracking only appears for the smallest value of $h$, whereas for the two largest values, there is only overturning. For $h=.7; .9$ neither cracking nor overturning occurs.

This behaviour  is confirmed in figure 9. Here we see how the strongest and deepest crackings are associated with  the smallest values of $h$. For some minimal value of $h$, and larger, only overturnings are observed. These are strongest for the largest values of $h$.

Figures 10 and 11 also confirm the role of $h$ in the occurrence (or not) of crackings. In figure 10, for $h=.5$, there are only crackings, which are strongest and deepest for the largest values of $\alpha$. In contrast, figure 11 only depicts overturnings for $h=1.5$ and different values of $\alpha$.

Figures 12, 13 and 14 describe the situation resulting from  the perturbation of $n$. However, unlike  the previous cases, for these three figures we have assumed negative values of $\Gamma$. This explains why they  only exhibit overturnings, and why these overturnings are stronger for smaller values of $h$ (just the opposite of the situation observed for positive values of $\Gamma$). 

\section{Conclusions}
We have investigated  the conditions under which,  general relativistic polytropes for anisotropic matter,  exhibit cracking and/or overturning, when submitted to fluctuations of energy density and anisotropy.

Thus, we have shown that cracking and/or overturning occur for a wide range of the parameters. For both types of polytropes, the main conclusions are basically the same, namely: the strongest and deepest crackings  occur for the smallest values of $h$ and greatest values of $\alpha$. Also, the strongest overturnings result from the largest values of $h$.

A distinct feature of all the models studied here, is the fact that the ``core'' (the inner part), and the ``envelope'' (the outer part),  respond differently to different degrees of anisotropy (different $h$). This fact was already pointed out in \cite{2p, HRW}.

It is important to stress that the occurrence of a cracking,  has direct implications on the structure and evolution of the compact object, only at time scales that are smaller than, or at most, equal to, the hydrostatic time scale.  This is so because as already mentioned, what we do is to take a ``snapshot'' just after the system leaves the equilibrium. To find out whether or not the system will return to the state of equilibrium afterward, is out of the scope of our analysis, and would require an integration of the evolution equations for a finite period of time, greater than the hydrostatic time.

However, all this having been said, it is clear that the occurrence of cracking would drastically affect the future structure and evolution of the compact object.

Accordingly,  we would like to conclude this work by speculating about possible
scenarios where the occurrence of cracking might be invoked, in order to understand the related observational data. 

One of these situations could be the
collapse of a supermassive star. The occurrence of  cracking at  the inner core, would
certainly change (in some cases probably enhance) the conditions for the
ejection of the outer mantle in a supernova event. This will be so for both
the ``prompt'' \cite{33, 34} and the ``long term'' mechanisms \cite{35, 36, 37, 38}.

Also, one is tempted to invoke cracking as the possible origin of quakes
in neutron stars \cite{39, 40,  41}. In fact, large scale crust cracking in neutron stars
and their relevance in the occurrence of glitches and bursts of x-rays and
gamma rays have been considered in detail by Ruderman (see \cite{42} and
references therein).

 Evidently, the characteristics of these quakes, and  those of the ensuing glitches, would strongly depend on the depth at which the cracking occurs. In this respect it is  worth noticing, the already mentioned fact, that  the depth at which the cracking may appear, is highly dependent on the parameter $h$, which measures the anisotropy of the pressure.

However we would like to emphasize that our aim
here is not to model in detail any of the scenarios mentioned above, but just 
to call  attention to the possible occurrence of cracking in such important configurations, as those satisfying a polytropic equation of state, and its relationship with fluctuations of local anisotropy. In other words, whatever  the origin of the anisotropy would be, or how small  it is, cracking may occur, which  would drastically affect  the outcome of the evolution of the system.


\end{document}